\documentclass{aa}

\usepackage[nolist,nohyperlinks]{acronym}

\usepackage{siunitx}
\usepackage{microtype}

\usepackage{enumitem}

\usepackage{widetext}

\usepackage{silence}
\usepackage[colorlinks, allcolors=blue, breaklinks=true]{hyperref}
\usepackage{cleveref}

\usepackage{xcolor}

\usepackage{amsmath}

\usepackage{changepage}

\usepackage{tabularx}

\usepackage{bm}

\usepackage{tikz}
\usetikzlibrary{shapes,arrows,fit,positioning,calc}

\let\endlongtable\relax

\usepackage[rightcaption, ragged]{sidecap}
\sidecaptionvpos{figure}{t}

\usepackage{tabularx}

\usepackage{txfonts}

\usepackage[encapsulated]{CJK}
\usepackage{ucs}

\newcommand{\orcid}[1]{\protect\href{https://orcid.org/#1}{\protect\includegraphics[width=8pt]{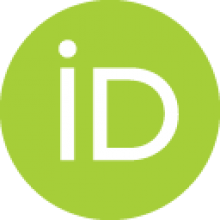}}}

\usepackage{xspace}

\defcitealias{eht-m87-paper-vii}{EHT et al. 2021a}

\def\M87{M87$^*$\xspace}
\def\m87{M87$^*$\xspace}
\def\sgra{Sgr~A$^*$\xspace}
\def\3C279{3C\,279\xspace}
\def\3c279{3C\,279\xspace}
\def\NRAO530{NRAO\,530\xspace}
\def\nrao530{NRAO\,530\xspace}
\def\J1924{J1924-2914\xspace}
\def\j1924{j1924-2914\xspace}
\def\z{\mbox{\textsc{Zingularity}}\xspace}
\def\symba{\mbox{\textsc{Symba}}\xspace}

\def\tensorflow{\mbox{\textsc{TensorFlow}}\xspace}
\def\tensorflowprobability{\mbox{\textsc{TensorFlow Probability}}\xspace}

\begin{document}

\begin{acronym}
\acro{sn}[S/N]{signal-to-noise ratio}
\acro{vlbi}[VLBI]{very long baseline interferometry}
\acroplural{vlbi}[VLBI]{Very-long-baseline interferometry}
\acro{grmhd}[GRMHD]{general relativistic magnetohydrodynamics}
\acro{grrt}[GRRT]{general relativistic ray-tracing}
\acro{mhd}[MHD]{magnetohydrodynamics}
\acro{as}[as]{arcseconds}
\acro{agn}[AGNs]{Active Galactic Nuclei}
\acro{llagn}[LLAGN]{low-luminosity AGN}
\acro{riaf}[RIAF]{radiatively inefficient accretion flow}
\acro{cena}[Cen~A]{Centaurus~A}
\acroplural{cena}[Cen~A, NGC~5128]{Centaurus~A}
\acro{sgra}[Sgr\,A*]{Sagittarius~A*}
\acro{m87}[\m87]{Messier 87*}
\acro{agn}[AGNs]{active galactic nuclei}
\acro{eht}[EHT]{Event Horizon Telescope}
\acro{bhc}[BHC]{\href{https://blackholecam.org}{BlackHoleCam}}
\acro{tanami}[TANAMI]{Tracking Active Galactic Nuclei with Austral Milliarcsecond Interferometry}
\acro{smbh}[SMBH]{supermassive black hole}
\acroplural{smbh}[SMBHs]{supermassive black holes}
\acro{aa}[ALMA]{Atacama Large Millimeter/submillimeter Array}
\acro{ap}[APEX]{Atacama Pathfinder Experiment}
\acro{pv}[PV]{IRAM~30\,m Telescope}
\acro{jc}[JCMT]{James Clerk Maxwell Telescope}
\acro{lm}[LMT]{Large Millimeter Telescope Alfonso Serrano}
\acro{sp}[SPT]{South Pole Telescope}
\acro{sm}[SMA]{Submillimeter Array}
\acro{az}[SMT]{Submillimeter Telescope}
\acro{c}[c]{speed of light}
\acro{pc}[pc]{parsec}
\acro{gr}[GR]{general relativity}
\acro{aips}[\textsc{aips}]{\href{http://www.aips.nrao.edu}{Astronomical Image Processing System}}
\acro{casa}[\textsc{casa}]{\href{https://casa.nrao.edu}{Common Astronomy Software Applications}}
\acro{symba}[\textsc{Symba}]{\href{https://bitbucket.org/M_Janssen/symba}{SYnthetic Measurement creator for long Baseline Arrays}}
\acro{rpicard}[\textsc{Rpicard}]{\href{https://bitbucket.org/M_Janssen/picard}{Radboud PIpeline for the Calibration of high Angular Resolution Data}}
\acro{hops}[\textsc{hops}]{\href{https://www.haystack.mit.edu/tech/vlbi/hops.html}{Haystack Observatory Postprocessing System}}
\acro{hbt}[HBT]{Hanbury Brown and Twiss}
\acro{tov}[TOV]{Tolman-Oppenheimer-Volkoff}
\acro{mri}[MRI]{magnetorotational instability}
\acro{adaf}[ADAF]{advection-dominated accretion flow}
\acro{adios}[ADIOS]{adiabatic inflow-outflow solution}
\acro{cdaf}[CDAF]{convection-dominated accretion flow}
\acro{bz}[BZ]{Blandford-Znajek}
\acro{bp}[BP]{Blandford-Payne}
\acro{em}[EM]{electromagnetic}
\acro{blr}[BLR]{broad-line region}
\acro{nlr}[NLR]{narrow-line region}
\acro{ism}[ISM]{interstellar medium}
\acro{edf}[eDF]{electron distribution function}
\acro{pic}[PIC]{particle-in-cell}
\acro{sane}[SANE]{standard and normal evolution}
\acro{mad}[MAD]{magnetically arrested disk}
\acro{jive}[JIVE]{\href{http://www.jive.eu}{Joint Institute for VLBI ERIC}}
\acro{nrao}[NRAO]{\href{https://www.nrao.edu}{National Radio Astronomy Observatory}}

\acro{muas}[$\mu$as]{microarcseconds}
\acroplural{agn}[AGNs]{Active galactic nuclei}
\acro{jy}[Jy]{Jansky}
\acro{pa}[PA]{position angle}
\acro{srmhd}[SRMHD]{special relativistic magnetohydrodynamics}
\end{acronym}

\title{Deep learning inference with the Event Horizon Telescope}
\subtitle{II. The \z{} framework for Bayesian artificial neural networks}

   \author{M. Janssen\orcid{0000-0001-8685-6544}
          \inst{1,2}
    \and C.-k. Chan\orcid{0000-0001-6337-6126}\inst{3,4,5}
    \and J. Davelaar\orcid{0000-0002-2685-2434} \inst{6,7}
    \and M.~Wielgus\orcid{0000-0002-8635-4242}\inst{8}
          }
    \institute{Department of Astrophysics, Institute for Mathematics, Astrophysics and Particle Physics (IMAPP), Radboud University, P.O. Box 9010, 6500 GL Nijmegen, The Netherlands\\
    \email{M.Janssen@astro.ru.nl}
   \and Max-Planck-Institut f\"ur Radioastronomie, Auf dem H\"ugel 69, D-53121 Bonn, Germany
            \and
            Steward Observatory and Department of Astronomy, University of Arizona, 933 N. Cherry Ave., Tucson, AZ 85721, USA
            \and
            Data Science Institute, University of Arizona, 1230 N. Cherry Ave., Tucson, AZ 85721, USA
            \and
            Program in Applied Mathematics, University of Arizona, 617 N. Santa Rita Ave., Tucson, AZ 85721, USA
            \and
            Department of Astrophysical Sciences, Peyton Hall, Princeton University, Princeton, NJ 08544, USA
            \and
            NASA Hubble Fellowship Program, Einstein Fellow
            \and
            Instituto de Astrofísica de Andalucía-CSIC, Glorieta de la Astronomía s/n, E-18008 Granada, Spain
             }

   \date{Received TBD; accepted TBD}

\nocite{eht-paperII}
\nocite{eht-paperIII}
\nocite{eht-paperIV}
\nocite{eht-paperV}
\nocite{eht-paperVI}
\nocite{2021eht-m87-mwl}
\nocite{eht-m87-paper-vii}
\nocite{eht-m87-paper-viii}
\nocite{eht-SgrAii}
\nocite{eht-SgrAiii}
\nocite{eht-SgrAiv}
\nocite{eht-SgrAv}
\nocite{eht-SgrAvi}

\abstract
{In this second paper in our publication series, we present the open-source \z{} framework for parameter inference with deep Bayesian artificial neural networks. We carried out out supervised learning with synthetic millimeter very long baseline interferometry observations of the Event Horizon Telescope (EHT). Our ground-truth models are based on general relativistic magnetohydrodynamic simulations of \sgra and \m87 on horizon scales. The models predict the synchrotron emission produced by these accreting supermassive black hole systems.}
{We investigated how well \z{} neural networks are able to infer key model parameters from EHT observations, such as the black hole spin and the magnetic state of the accretion disk, when uncertainties in the data are accurately taken into account.}
{\z{} makes use of the \tensorflowprobability{} library and is able to handle large amounts of data with a combination of the efficient \texttt{TFRecord} data format plus the \textsc{Horovod} framework for distributed deep learning. Our approach is the first analysis of EHT data with Bayesian neural networks, where an unprecedented training data size, under consideration of a closely modeled EHT signal path, and the full information content of the observational data are used. \z{} infers parameters based on salient features in the data and is containerized for scientific reproducibility.}
{Through parameter surveys and dedicated validation tests, we identified neural network architectures, that are robust against internal stochastic processes and unaffected by noise in the observational and model data. We give examples of how different data properties affect the network training. We show how the Bayesian nature of our networks gives trustworthy uncertainties and uncovers failure modes for uncharacterizable data.}
{It is easy to achieve low validation errors during training on synthetic data with neural networks, particularly when the forward modeling is too simplified. Through careful studies, we demonstrate that our trained networks can generalize well so that reliable results can be obtained from observational data.}


   \maketitle

\ActivateWarningFilters[hreflink]
\section{Introduction}
\label{sec:intro}

Low-powered \ac{agn} are well described by numerical \ac{grmhd} simulations \citep[e.g.,][]{2003DeVilliers, 2006McKinney, 2012Dibi, 2018Ryan}. \ac{grmhd} simulations solve the equations of a magnetohydrodynamic fluid in curved spacetime. An initial setup with a weakly magnetized torus self-consistently evolves into radiatively inefficient accretion flows accompanied by outflows and jets.

In this work, we focus on the low-luminosity \ac{agn} \ac{sgra} and \ac{m87}. \m87 is a nearby elliptical galaxy in the Virgo cluster and features a prominent extragalactic radio jet that has been resolved by observations in the radio to X-ray bands \citep[e.g.,][]{Curtis1918, 1966Byram, 1989Owen, 1996Sparks, 1999Biretta, 2002Marshall, Hada2011, 2016Mertens, Walker2018}. \sgra is known as the ``starving'' supermassive black hole in our Galactic Center with a very low accretion rate and no visible jet emission, discovered as a bright radio source \citep{1974Balick}. Detections of a gravitational redshift \citep{2018GRAVITYredshift, 2019Do} and Schwarzschild precession \citep{2020GRAVITY, 2022Gravity} of a star in orbit around the black hole served as important tests of General Relativity and for over two decades \sgra's rich multiwavelength variability has been studied. Here, we refer to the detailed recent studies by \citet{2021Witzel} and \citet{eht-SgrAii} and references therein.

Electromagnetic observables computed from \ac{grmhd} models through \ac{grrt} are overall in good agreement with high-resolution radio observations of \ac{sgra} and \ac{m87}.
Selections of a ``standard'' set of \ac{grmhd}-\ac{grrt} models (\citet{eht-paperV, eht-SgrAv}; see also the overview of the model space given in \citet{zingularity1}) have been scored against high-resolution millimeter-wave observations of \ac{sgra} and \ac{m87} with the \ac{eht} \ac{vlbi} array in various ways.

In \citet{eht-paperV}, a $\chi^2$ scoring of 100-500 snapshots per \ac{m87} model against the \ac{eht} total intensity data of \ac{m87} with the \textsc{Themis} \citep{themis} and \textsc{Gena} \citep{gena} software packages disfavored \ac{mad} accretion models, which have strong and organized magnetic fields that can disrupt the accretion flow \citep[e.g.,][]{2012McKinney, 2012Narayan} and highly negative black hole spins (measured with respect to the accretion flow). For this average image scoring analysis, closure phases \citep{1958Jennison} and total intensity amplitudes were used. The measurements were averaged over minute-long \ac{vlbi} scan, which can lead to decoherence on low \ac{sn} data, where atmospheric phase turbulence cannot be calibrated reliably. 

In \citet{eht-m87-paper-viii}, the image-integrated linear and circular polarization, average linear polarization, and a parameter describing the azimuthal electric-vector position angle pattern \citep{Palumbo2020, 2022Palumbo, 2022Emami} derived from the \ac{m87} linear polarization maps at 230\,GHz \citep{eht-m87-paper-vii, 2021Goddi} have been compared against a 200 \ac{grrt} image frames subset blurred with a $20\,\mu$as beam for each model. Overall, MAD models are in better agreement than \ac{sane} models, where the magnetic fields are weaker and more turbulent. The MAD preference is reinforced when the measured upper limits on resolved circular polarization are taken into account \citep{2023EHTStokesV}.

In \citet{eht-SgrAv}, parts of the \ac{eht} \sgra data were used for source-structural constraints based on the image size, salient features in the measured flux densities, and geometric ring-model fits \citep{eht-SgrAiv}, making use of the new \textsc{Comrade} software \citep{2022Comrade}, among others. Models with negative spin, edge-on inclination angles, and equal ion and electron temperatures are disfavored.

The 2017 \ac{eht} \sgra polarization data were analyzed in \citet{2024eht-sgra-pol}. The \mbox{\textsc{KerrBAM}} semianalytic model \citep{KerrBAM} was used for a qualitative exploration of the relations between polarization measurements and physical quantities of \sgra. The \ac{grmhd} scoring leaves a single passing MAD model at an inclination of \ang{150} with a spin of 0.94 and comparatively strong jet emission, assuming that the observed rotation measure is produced by an external Faraday screen.

Multiwavelength observations of \ac{m87} \citep{2021eht-m87-mwl} rule out spin-zero models (based on jet power) and SANE models, where very hot electrons in the accretion disk produce an X-ray luminosity excess \citep{eht-paperV}.
\ac{sgra} multiwavelength data \citep{eht-SgrAii} favors MAD models and rejects most models with large inclination angles and all models with an equal ion and electron temperature. The majority of the models are more variable than the observed intra-day total-flux light-curve flux variations \citep{eht-SgrA_LC}, while SANE models are overall less variable than MADs \citep[see the discussion in][]{eht-SgrAv}.

Using one specific image feature, the size of the black hole shadow \citep{2000Falcke} calibrated with \ac{grmhd}-\ac{grrt} models, \citet{2020Psaltis}, \citet{2021Kocherlakota}, and \citet{eht-SgrAvi} performed tests of spacetime metrics.
The degree to which the shadow can be used for a clean test of gravity through a robust relation to the photon ring \citep[e.g.,][]{1973Bardeen_phorb, 2020Johnson} is being debated and depends on how well we understand low-powered AGN accretion physics \citep{2019Narayan, 2021Gralla, 2021Chael, 2021Bronzwaer, 2021arXiv211101123O, Wielgus2021, 2022Vincent, 2022Paugnat}.

Recently, further \ac{eht} data analysis methods have been proposed. \citet{2023MedeirosA, 2023MedeirosB} used dictionary learning of a \ac{grmhd}-\ac{grrt} simulations to reconstruct images from \ac{eht} data through a principal components analysis.
\citet{2023Emami} have shown that $E$ and $B$ linear polarization modes are informative regarding \ac{grmhd}-\ac{grrt} parameters.
\citet{2023Conroy} have devised a method to measure rotation speeds in future \ac{eht} movie reconstructions using autocorrelations in the image domain.
\citet{2023Chael} have identified a relation between horizon-scale polarimetric observables and electromagnetic energy extraction from the black hole spin.
\citet{2024Yfantis} have developed a Bayesian ray-tracing parameter estimation framework, related to earlier work by \citet{2016Kim} and \citet{2022Psaltis}.
New methods for directly reconstructing the source structure for analyses in the image domain have been presented in \citet{2022Mueller}, \citet{2023Mueller}, \citet{2024bMus}, \citet{2024Mus}, and \citet{2024aMus}.

In this work we present \z,\footnote{\url{https://gitlab.com/mjanssen2308/zingularity}} an open-source generic \tensorflow{}-based \citep{tf1, tf2} framework of Bayesian deep neural networks for astronomical interferometers, which we developed to use the full information content of data from \ac{eht} observations for a \ac{grmhd}-\ac{grrt} data-driven parameter inference without the need to compromise on the number of model images due to computational limitations.
This is realized by using large-scale synthetic data libraries that are based on a wide range of \ac{sgra} and \ac{m87} simulations \citep{zingularity1}.
Here we describe and validate the \z{} network training for the 2017 \ac{eht} observations.
\citet{zingularity3} shows the parameter inference results from the application of the trained network to observational data alongside predictions of future upgraded \ac{eht} observations using models that go beyond the \citet{Kerr_1963} metric.

Driven by the vastly increasing data sizes, machine learning methods are becoming increasingly popular in astronomy, mostly in the area of source finding as well as classification \citep[see for example][]{2012cidu.conf...47V, 2019arXiv190407248B, 2020FlukeJacobs, 2020bda..book.....K, 2022arXiv221103796S, 2022Djorgovski, 2023PASA...40....1H, 2023Moriwaki} and with convolutional neural networks \citep[e.g.,][]{2015Natur.521..436L} in particular.
Related to our work, \citet{2019JETP..128..592S}, \citet{eht-ml1}, \citet{eht-ml2}, and \citet{2021A&C....3600467P} have used neural networks in the \ac{eht} total intensity image domain, where the uncertainties and model degeneracies are large for the current \ac{eht} \mbox{($u$, $v$)} coverage.
\citet{2023Qiu} have trained a random forest machine learning model on a few salient features in the image domain from \ac{grmhd}-\ac{grrt} simulations, including polarization information. \citet{Levis2024} used neural networks to study the orbital dynamics around \ac{sgra} in the millimeter wavelength polarimetric observations.
Compared to \citet{eht-ml3}, where total intensity visibility measurements from \ac{m87} were used for the network training, we used a larger training dataset for \ac{m87}, also considered  the \ac{sgra} data, included more effects for the modeling of the \ac{eht} signal path, took the additional polarization information from the measurements into account, and obtained uncertainties on the inferred parameters by using a Bayesian network and bootstrapping of observational data errors.

In terms of more general machine learning applications for astronomical interferometers, \citet{2020arXiv200310424S} have developed a method for optimizing the telescope locations for \ac{vlbi} arrays, \citet{2020arXiv201014462S} present a variational deep probabilistic imaging approach, quantifying reconstruction uncertainty, and \citet{2022ApJ...932...99S} use an improved variational inference method to obtain accurate posterior samples for parameter inference tasks.
Neural-network-based interferometric imaging methods have been developed by \citet{2018arXiv180800011M, 2019ApJ...883...14M}, \citet{2022Gheller}, \citet{2022arXiv220311757S}, \citet{2024Aghabiglou}, \citet{2024Thyagarajan}, \citet{2024Feng}, and \citet{2024Lai}.
\citet{2024Mohan} and \citet{2024SaraerToosi} use neural networks to generate \m87 images.
Finally, \citet{2022MNRAS.512.5848D} have used a deep convolutional neural network as a fast generator of \ac{grmhd} simulations, a method that might substantially speed up the generation of synthetic training data in the future.

In \Cref{sec:concepts} of this work we introduce machine learning and astrophysical interferometry concepts used throughout this paper.
In \Cref{sec:zingularitymotivation}, we give our motivation for developing \z{} (i.e., the usefulness of ML applications for the analysis of data from astronomical interferometers like the EHT).
In \Cref{sec:zingularitytrainingdata}, we introduce the \ac{grmhd}-\ac{grrt} synthetic data library of \sgra and \m87 EHT observations used as a training dataset for \z{} in this work.
In \Cref{sec:zingularity}, we describe the \z{} framework together with the specific algorithms used to fit models to EHT data in this paper.
In \Cref{sec:zingularity-validation}, we show the suite of validation tests and diagnostics implemented to verify the output of \z{}, particularly under the aspect of how \z{} performs Bayesian model parameter inference from uncharacterized data and how much this inference is hindered by instrumental effects.
We offer our conclusions about \z{} and the ability to extract \ac{grmhd}-\ac{grrt} parameters from current EHT observations in \Cref{sec:zingularityconclusions}.
We finish with an outlook of planned future studies with \z{} in \Cref{sec:zingularityoutlook}.

\section{Machine learning and interferometry concepts}
\label{sec:concepts}

Machine learning (ML) methods automatically learn the characteristics of a training dataset $\widetilde{T}$.
In this work, supervised learning is carried out, where labeled training data are used. As such, the ML algorithm optimizes itself based on a known input-output mapping.
We are using a fraction of the data contained in $\widetilde{T}$ as validation data, which is used to compute the accuracy of the trained ML model based on data that was not used for the optimization.

Artificial neural networks (ANNs) are ML algorithms that are designed based on neural connections of biological brains.
Information in the form of floating point numbers are passed from an input dataset, through connected computing nodes (`neurons') that each transform the data with some function $g$, up to a final layer, that yields the predictions of the ANN.
A network organized in layers, for which the same type of operation $g(x)$ is computed for each neuron and the input $x$ is taken as the output of the previous layer in the case of feedforward networks (\Cref{sec:NNbackground}).
Every network has one input and one output layer. ANNs with more than three layers are referred to as deep neural networks.
Free parameters of each neuron's $g$ are optimized (`trained') based on a loss function between the network's input and output.

We refer to all data imperfections (i.e., all differences in the data measured between a realistic instrument and a hypothetical perfect measurement device) as data corruption effects along the signal path $\widetilde{C}$.
These effects encompass thermal noise and uncorrected systematics from the instrument itself, corruptions that occur along the long signal path of astronomical observatories (e.g., interstellar scattering and the added noise and absorption from Earth's atmosphere for ground-based observatories) as well as uncorrected systematics introduced by assumptions made during the data reduction process.

We denote uncharacterized data, for which we do not know the underlying physical reality as $\widetilde{U}$. Typically, $\widetilde{U}$ is obtained from a measurement and affected by $\widetilde{C}$. We assume $\widetilde{U}$ to be reasonably well described by a model $\widetilde{M}$ and wish to infer the model parameters from the data.
For our EHT horizon-scale observations of \sgra{} and \m87{}, we use \ac{grmhd}-\ac{grrt} images as $\widetilde{M}$.

With synthetic data $\widetilde{S}$, we attempt to create mock observations that sample the data produced by an astronomical observatory as closely as possible.
This requires a full forward modeling, where the physical reality is well described by $\widetilde{M}$ and all relevant data corruption effects along the signal paths are modeled correctly.
We used synthetic data based on \ac{grmhd}-\ac{grrt}  images for our training input of our ANN as described in \citet{zingularity1}: $\widetilde{T} = \widetilde{S}(\widetilde{M}, \widetilde{C})$.
It is therefore required that the most significant features of $\widetilde{S}$, that have discriminative power for $\widetilde{M}$ in the presence of $\widetilde{C}$, to also be present in $\widetilde{U}$. \ac{grrt} images are currently our best models for the interpretation of EHT observations \citep{eht-paperV, PaperV}. We note that that restriction to \ac{grmhd}-\ac{grrt} makes our analyses model-dependent and that the validity of these models is also being questioned in the literature \citep[e.g.,][]{2021Gralla}.
While we make use of a large parameter space of the used models, all simulations assume ideal MHD and a simple description of the electron temperature as described in the previous paper in this series. Further limitations of the models used in this initial study are described in the outlook \cref{sec:zingularityoutlook}.

Most astronomical instruments measure the electric field emitted by a radiating source with sky brightness distribution $I$.
Interferometers cross-correlate the signals measured by pairs of telescopes to create visibilities $\mathcal{V}$.
A Fourier relationship links $I$ and uncorrupted $\mathcal{V}$ \citep{1934vanCittert, 1938Zernike}. However, interferometers provide only an incomplete sampling of $\mathcal{V}$ corresponding to projected vectors connecting all pairs of telescopes (baselines).
Hence, obtaining $I$ from $\mathcal{V}$ becomes an ill-posed problem and additional information or assumptions must be used \citep{2017Thompson}.
Furthermore, $\widetilde{C}$ must be modeled to correct the $\mathcal{V}$ measurements \citep{1996Hamaker}.
Visibilities are measured as a function of time $t$, frequency $\nu$, telescope baseline vector $(u, v, w)$, and polarization $P$.
Each telescope in the interferometer typically measures two orthogonal polarization states of the radiation (right/left-handed circular polarization or horizontal/vertical linear polarization).
Four polarization products $P$ are formed from the two telescopes that form a baseline and the two orthogonal polarization measurements.
The four Stokes parameters \citep{1851Stokes} can be formed from different linear combinations of the four $P$ values.

ML methods build a complex model that tries to capture the most important features of $\widetilde{M}$ through a training dataset $\widetilde{T}$, such that parameters of $\widetilde{M}$ can be retrieved from $\widetilde{U}$ indirectly.
We will be referring to methods that are fitting $\widetilde{M}$ directly to $\widetilde{U}$ as conventional. For conventional interferometer methods, $\widetilde{M}$ are typically simple geometric models or pixel-based images $I$. The direct fitting of complex models such as \ac{grrt} images poses a significant computational challenge. We summarize our shorthand for the machine learning concepts in \Cref{tab:terminology}.

\begin{table}[t]
\caption{Shorthand for  terminology used in this work.}
\begin{center}
\begin{tabularx}{\linewidth}{l l}
\hline
\hline
Symbol & Description \\
\hline
\rule[0pt]{0pt}{\heightof{A}+1ex} $\widetilde{U}$ & Uncharacterized data; here the observational data \\
\rule[0pt]{0pt}{\heightof{A}+1ex} $\widetilde{C}$ & Data corruption effects along the signal path \\
\rule[0pt]{0pt}{\heightof{A}+1ex} $\widetilde{M}$ & Model data \\
\rule[0pt]{0pt}{\heightof{A}+1ex} $\widetilde{S}$ & Synthetic data \\
\rule[0pt]{0pt}{\heightof{A}+1ex} $\widetilde{T}$ & Training data; here $\widetilde{T} = \widetilde{S}(\widetilde{M}, \widetilde{C})$ \\
\hline
\end{tabularx}
\label{tab:terminology}
\end{center}
\end{table}

\section{Motivation}
\label{sec:zingularitymotivation}

\begin{table*}[t]
\caption{\z training and application parameters. For \sgra, we found two equally viable models with different $N_\mathrm{ep}$.}
\begin{center}
\begin{tabularx}{\linewidth}{@{\extracolsep{\fill}}l|clll}
\hline
\hline
\hline 
Category  & Param.  &  \multicolumn{2}{l}{\;\;\;\;\;\;\;\;\;\;Value} & Description \\
\hline
\hline 
Common         & $f$ & \multicolumn{2}{l}{Swish \citep{2017arXiv171005941R}} & Activation function for all hidden layers \\
\z{}         & $\Xi$ & \multicolumn{2}{l}{RMSProp} & Optimization algorithm \\
GRMHD     & $L$ & \multicolumn{2}{l}{Negative log-likelihood}  & Loss function\\
-GRRT EHT    & $\eta_\mathrm{val}$ & \multicolumn{2}{l}{$0.1$} & Fraction of $\widetilde{T}$ used for validation \\
parameters   & $N_\mathrm{b}$ & \multicolumn{2}{l}{256} & Training batch size \\
       & $l_r$ & \multicolumn{2}{l}{$0.001 \times n / N_\mathrm{ep}$} & Learning rate warm-up for $0 \leq n \leq 0.1 \times N_\mathrm{ep}$ \\
        &       & \multicolumn{2}{l}{$0.0001 \times ( 1 + \cos(n \pi/N_\mathrm{ep}) ) / 2$} & Learning rate cosine decay for $0.1 \times N_\mathrm{ep} \leq n \leq N_\mathrm{ep}$ \\
\hline
        & & \bf{\m87} & \bf{\sgra} & \\
Fiducial        & $N_\mathrm{vis}$ & $8\times5489$ & $8\times13840$ & Number of data points in a training sample \\
models           & $N_\mathrm{tr}$ & 600,000 & 252,000 & Number of training samples \\
                & $N_\mathrm{ep}$ & 70 & 50, 60 & Number of training epochs \\
                & $\eta_\mathrm{drop}$ & $0.01$ & 0 & Dropout rate for stochastic neuron deactivation \\
                & $\mathcal{L}_1$ & $0.01$ & 0.01 & $L_1$ (lasso) regularization hyperparameter\\
                & $\mathcal{L}_2$ & $0.01$ & 0.01 & $L_2$ (ridge) regularization hyperparameter\\
                & $k_\mathrm{conv}$ & 8  & 8  & Receptive field of CNN layers \\
                & $n_\mathrm{CNNb}$ & 16  & 8  & Baseline number of neurons for the ResNet CNN layers \\
                & $n_\mathrm{CNNl}$ & 128  & 2048  & Neurons in last ResNet CNN layer \\
                & $N_\mathrm{dense}$ & 15  & 12    & Number of post-ResNet dense variational layers \\
                & $n_\mathrm{dense}$ & 128 & 1024  & Neurons in post-ResNet dense variational layers \\
                & $N_\mathrm{free}$ & 1,376,806 & 135,068,877 & Number of free parameters in the network \\
\hline
Boot-                & $\mathcal{D}$ &  \multicolumn{2}{l}{1\,--\,3\,\% \citepalias{eht-m87-paper-vii}}  & Polarization leakage ($\mathcal{D}$-terms) \\
strapping          & $\mathcal{G}_\mathrm{planet}$ &  \multicolumn{2}{l}{10\,\% \citep{Janssen2019b}} & Primary calibrator model uncertainties  \\
errors for              & $\mathcal{G}_\mathrm{scatter}$ &  \multicolumn{2}{l}{5\,--\,35\,\% \citep{Janssen2019b}} & DPFU uncertainty due to measurement scatter \\
the EHT                     & $\mathrm{gc}_\mathrm{B}$ &  \multicolumn{2}{l}{3.6\,--\,10.4\,\% \citep{Janssen2019b}} & Measurement error on gain curve curvature \\
                     & $\mathrm{gc}_\mathrm{E0}$ &  \multicolumn{2}{l}{1\,--\,2\,\% \citep{Janssen2019b}} & Measurement error on gain curve peak elevation \\
                     & $\sigma_\mathrm{th}$ &  \multicolumn{2}{l}{$\sim 8.5\times 10^{-6} \sqrt{\mathrm{SEFD}_1\mathrm{SEFD}_2}$} & Thermal noise of \ac{eht} data used in this work\\
\hline
\hline
\end{tabularx}
\label{tab:zingularity}
\end{center}

\end{table*}

In this section, we summarize the motivation for developing \z{}, which are the use-cases and unique opportunities when employing ML for data from astronomical interferometers and the \ac{eht} in particular.

Firstly, the \tensorflow{} library makes efficient use of the modern \texttt{TFRecord}\footnote{\url{https://www.tensorflow.org/tutorials/load_data/tfrecord}.} data format and scales well with CPU, GPU, and TPU\footnote{\url{https://cloud.google.com/tpu/docs/tpus}} computing power.
This allows \z{} to be trained on large $\widetilde{T}$ that span a wide range of $\widetilde{M}$ and $\widetilde{C}$ parameters. For the \ac{eht}, \z{} can make inferences based on the full \ac{grmhd}-\ac{grrt} parameter space, which is not computationally feasible with conventional methods. Furthermore, \z{} is able to process visibilities to their full extend, without losses that can occur when data are being averaged.
Once the network is trained, the application to $\widetilde{U}$ for inference is practically instantaneous.
Efficient and scalable analysis software will be needed for the increasing data rates of future instruments. In particular, the \ac{eht} plus next-generation \ac{eht} \citep{2019BlackburnngEHT,2023Johnson,2023Doeleman} will observe with an increased bandwidth as well as more telescopes in the future \citep{ehtmidscience} and exascale computing will be needed to handle the data produced by the Square Kilometre Array \citep{Dewdney2009} and next-generation Very Large Array \citep{2019ngVLA}. It has already been demonstrated that such big computing tasks are achievable with \tensorflow{} \citep{tf-exascale}.
More details about the computational speed of \z{} are given in \Cref{sec:benchmark-tests}.

Secondly, supervised ML is designed to obtain results with well defined fidelity metrics that describe how accurately model parameters can be recovered in a traceable manner (\Cref{sec:metrics}).

Thirdly, ANNs are training on salient and robust features by design.
As such, data points that are strongly affected by $\widetilde{C}$ or that are not being distinctive for the underlying $\widetilde{M}$ parameters of interest are not taken into account.
In simple $\chi^2$ analyses for the goodness of fit, these data points can lead to poor results.
Closure phases \citep{1958Jennison}, log closure amplitudes \citep{2020Blackburn}, and closure traces \citep{2020Broderick} are known robust observables that can be formed from visibilities, but whose errors are no longer Gaussian in the low \ac{sn} regime and whose variance does depend on telescope gain errors \citep{2022Lockhart}.
GRMHD scoring employed in past EHT analyses is dependent on how the method is implemented as well as how observational and model uncertainties are treated.
Deep ANNs may uncover more complex robust data combinations, which are discriminative for the $\widetilde{M}$ parameter space. Especially for complex models, such as GRRT images, the parameter-to-feature correspondence is not fully known a priori.

Fourthly, by using large $\widetilde{T}$ consisting of realistic synthetic data, all forward-modeled data corruption effects are taken into account by ML methods.
For sparse interferometric measurements, data corruptions \citep[e.g., antenna gain errors, instrumental polarization leakage, and atmospheric effects for high frequency observations,][]{zingularity1} can have a substantial influence on the obtained results.
The modeling of these (time-variable) effects is a) convoluted with the reconstructions of (time-variable) source structures during self-calibration \citep{Readhead1978,Pearson1984}, b) limited by the \ac{sn} of the observational data \citep[e.g.,][]{2022Janssen}, c) often based on simplifying assumptions \citep{eht-paperIII}, and d) introduces additional noise from the precision of determined calibration solutions \citep[e.g., fringe-fitting phase, delay, and rate estimates;][]{2017Thompson}.
Some algorithms can marginalize over a range of some $\widetilde{C}$ \citep[e.g.,][]{themis, dmc}, which requires considerable computational resources.
Moreover, a realistic modeling of the data acquisition and processing of $\widetilde{U}$ for the generation of $\widetilde{S}$ ensures that the results are not affected by unknown systematics introduced by a specific data reduction procedure.
\ac{vlbi}-specific data processing methods are often complex and the consequences of some calibration assumptions are not always fully explored and understood. One example is the usage of a point source during the fringe-fititng stage \citep{2020Natarajan}.
In analyses published by the \ac{eht} collaboration, a small percentage of systematic noise is added to deal with unknown data imperfections \citep{eht-paperIII, eht-SgrAii}. Here, we assert this to be covered by our forward modeling plus error bootstrapping. The flat addition of a single systematic noise budget to all baselines can lead to source signals being washed out on baselines that measure high correlated flux densities.

Fifthly, ML can easily be used to study the predictive power of $\widetilde{M}$ separately from $\widetilde{C}$, if the signal path and instrument can be modeled accurately.
Synthetic training data can be flexibly generated with different combinations of $\widetilde{C}$ and the corresponding accuracy of $\widetilde{M}$ parameter predictions can be studied.
The predictive power of $\widetilde{M}$ alone can be studied in the limit of a perfect hypothetical instrument, where no $\widetilde{C}$ is added to the synthetic data.
\citet{2020Roelofs} performed a simple image-based fidelity study of how well an upgraded \ac{eht} would be able to detect the M87 jet for example.
Another example, where the model predictions are straightforward, are self-similar photon-subrings surrounding the shadow of a black hole. These rings can be studied to test GR and to accurately measure black hole parameters \citep{2020Johnson}.
It is expected for the photon-ring signals to dominate at long interferometric baselines. With ML, one could for instance study the accuracy of a space-\ac{vlbi} GR test, marginalized over a range of possible accretion environments around the black hole, as a function of maximally achieved baseline length and calibration uncertainty.

Sixthly, \z{} makes use of the full information content of $\widetilde{T}$ and $\widetilde{U}$. Conventional mm \ac{vlbi} methods on the other hand often analyze the visibilities in stages: Using only the total intensity (Stokes $\mathcal{I}$) information first, followed by linear polarization (Stokes $\mathcal{Q}$ and $\mathcal{U}$) in some cases. The usually negligible circular polarization (Stokes $\mathcal{V}$) signals from the source and temporal evolution of $I$ are also often modeled separately, if at all.

Finally, while previous ANN-based parameter estimation studies for the \ac{eht} were based on Stokes $\mathcal{I}$ images \citep{eht-ml1, eht-ml2}, \z{} works with the visibilities directly and makes use of the full polarization information.
The advantages of using the visibilities directly are that the intermediate modeling step necessary to obtain $I$ is removed and that there is a precise description of $\mathcal{V}$ uncertainties.
Additionally, there are no constraints for image-specific parameters. Thus, different models, for example with different fields of view and pixel sizes, can be used together.

Before proceeding, the limitations of our current ML approach should be discussed as well. Like in the EHT GRMHD scoring, results have to be interpreted within the $\widetilde{M}$ parameter space.
Further, with our current BANN implementations, we are tied to specific \mbox{($u$, $v$)} coverages from specific observations.
Here, it is important to note that a proper forward modeling for unbiased inference requires the $\widetilde{T}$ generation to be tailored to the characteristics of specific observations anyway. We thus argue that computational efficiency could be gained by speeding up our $\widetilde{S}$ generation methods, rather than attempting to devise a flexible BANN that can be applied to observations it was not trained on.

\section{Training data}
\label{sec:zingularitytrainingdata}

\begin{figure*}
    \centering\offinterlineskip
    \includegraphics[height=13.682cm]{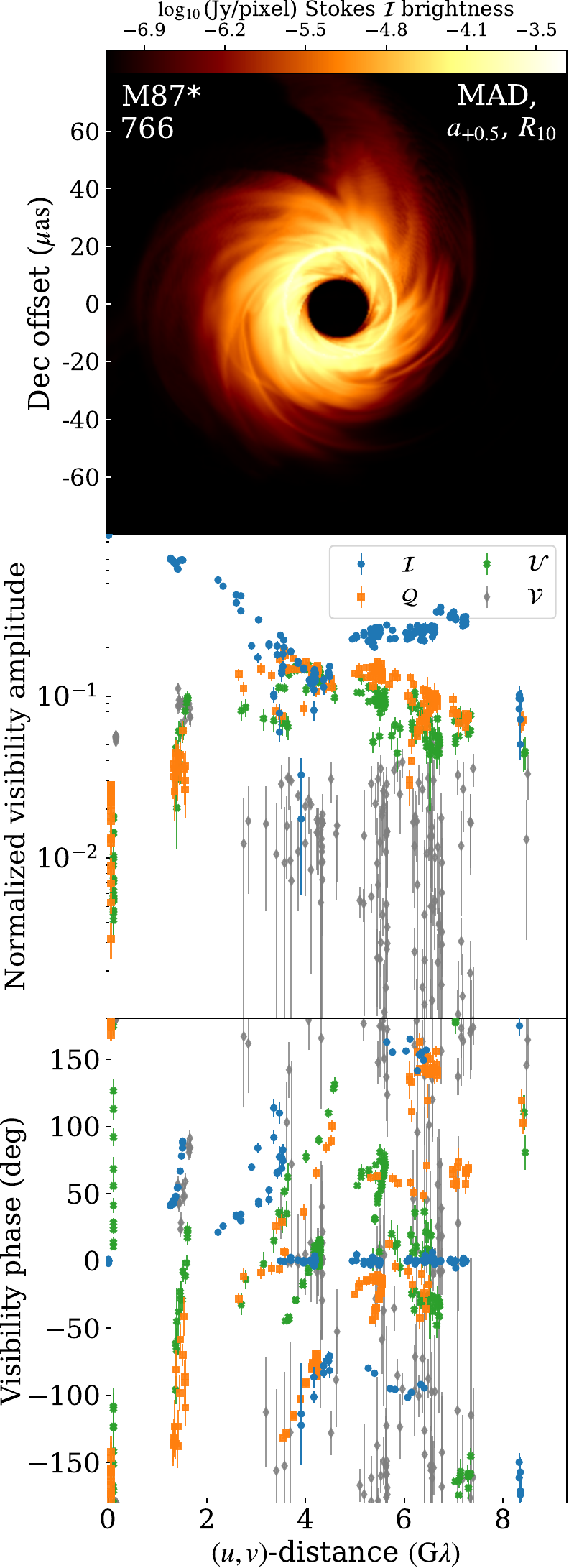}
    \hskip -0.6ex\includegraphics[height=13.682cm]{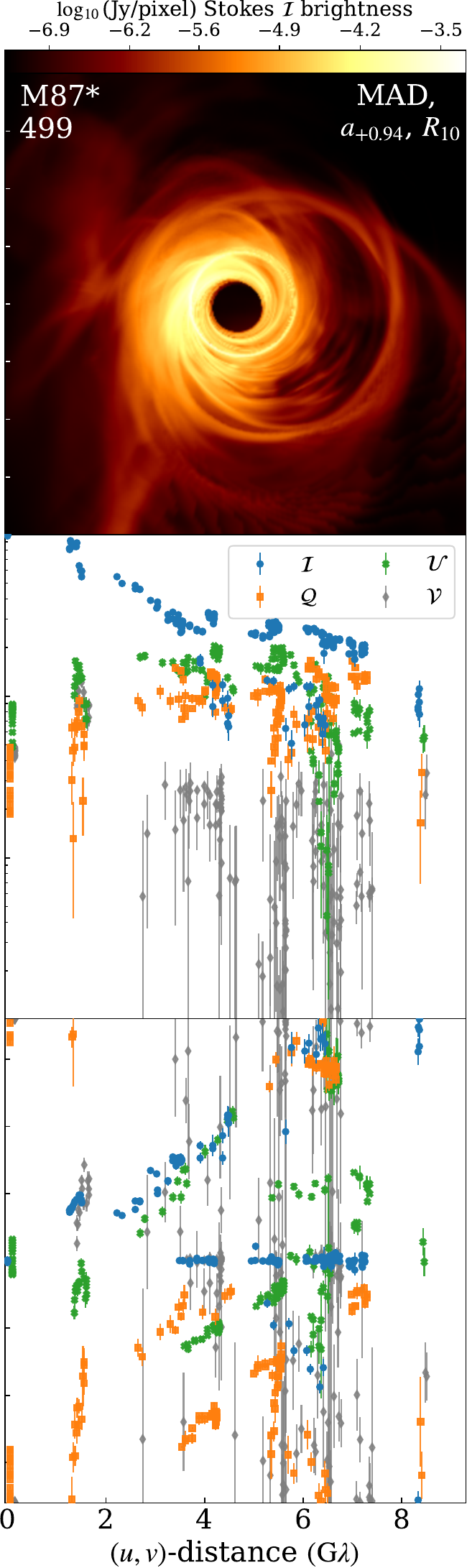}
    \hskip -0.6ex\includegraphics[height=13.682cm]{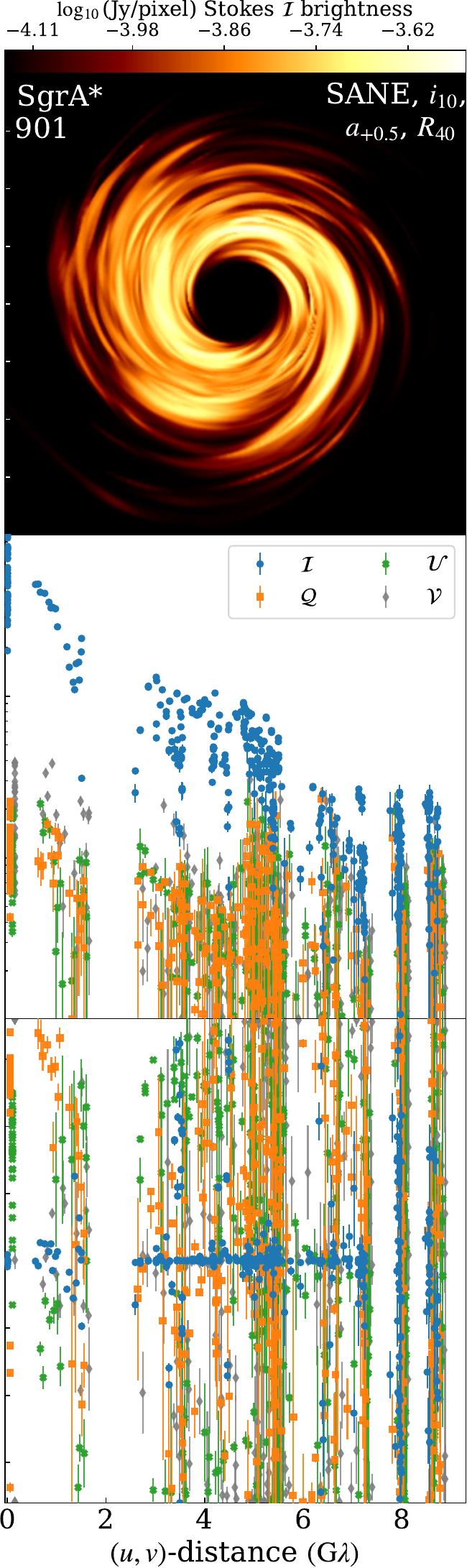}
    \hskip -0.6ex\includegraphics[height=13.682cm]{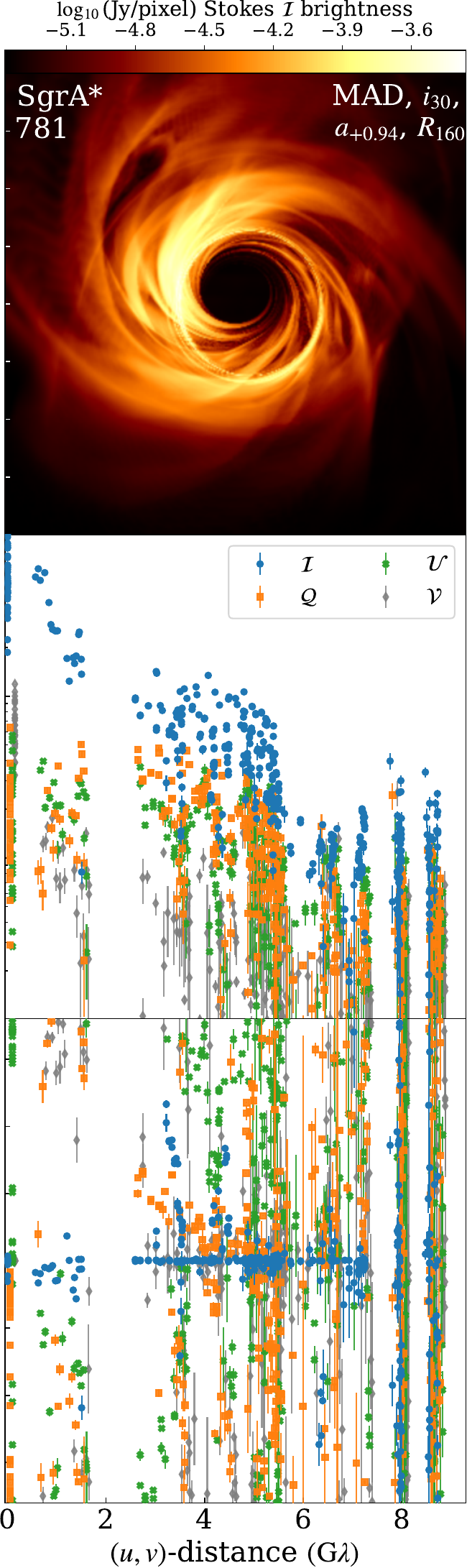}
    \caption{Four training dataset examples. The top row shows the total intensity ray-traced ground-truth model images on logarithmic scales with varying dynamic ranges. Normalized full-pol visibility amplitudes and phases of corresponding synthetic data realizations are displayed with thermal noise error bars as a function of baseline length in units of the observing wavelength $\lambda \approx 1.3$\,mm (see \citet{zingularity1} for the $(u, v)$ coverage) in the middle and bottom rows, respectively. The measurements shown can come from different orientations at the same baseline length. For a better readability, the visibilities have been averaged over scan durations, normalized amplitudes lower than 0.001 have been clipped, and the values of the different Stokes parameters are each offset by $50\,\mathrm{M}\lambda$ on the x-axis. In the top left corner of each model image is indicated whether the model and data correspond to \m87 or \sgra and the GRRT frame number of the image. The strongly time-variable \sgra data were generated from multiple GRRT frames, of which a single frame is displayed here. Spin $a_*=s$, $R_\mathrm{high}=r$, and $i_\mathrm{los}=l$ parameters are listed in a shorthand notation as $a_s$, $R_r$, and $i_l$ in the top right corner of each model image. The \sgra models are shown here with $\theta_\mathrm{PA}=0$.}
    \label{fig:training_data}
\end{figure*}

\begin{figure*}
\centering
\scalebox{.7}{
\tikzset{
    between/.style args={#1 and #2}{
         at = ($(#1)!0.5!(#2)$)
    }
}

\tikzstyle{action4} = [regular polygon,regular polygon sides=7, draw, fill=gray!20, 
    text width=5em, text centered,  minimum height=4em, node distance = 0.5cm]
\tikzstyle{action} = [draw, rectangle,fill=green!20, node distance=1cm,
    minimum height=2em, text width=8em, text centered]
\tikzstyle{action3} = [draw, ellipse,fill=orange!20, node distance=1cm,
    minimum height=2em, text width=6em, text centered]
\tikzstyle{input} = [rectangle, draw, fill=purple!20, 
    text width=8em, text centered, node distance=1cm and 7.5cm, minimum height =4em, rounded corners]
\tikzstyle{line} = [draw, -triangle 45]    

\begin{tikzpicture}[node distance = 2cm, auto]
    \node [input] (grmhd) {Input models (GRMHD)};
    \node [action, below = of grmhd] (grrt) {Ray-tracing};
    \node [input, right = of grmhd] (obs) {EHT observations};
    \node [action, below = of obs] (corr) {Correlation};
    \node [action, between=grrt and corr] (vlbimon) {\textsc{vlbimonitor}};
    \node [action, below = of grrt] (meqsil) {\textsc{MeqSilhouette}};
    \node[action, below = of meqsil] (rpicardsim) {\textsc{Rpicard}};
    \node[draw,dotted, inner sep=2mm,label= center:\underline{\textsc{symba}},fit=(meqsil) (rpicardsim)] (symba) {};
    \node[action, below = of rpicardsim] (osg) {Open Science Grid};
    \node[action3, below = of vlbimon] (rpicardparams) {Common processing parameters};
    \node[action3, below = of rpicardparams] (apcalmemo) {Station sensitivities \& uncertainties};
    \node[action, below = of corr] (rpicard) {\textsc{Rpicard}};
    \node[action, below = of rpicard] (fluxcal) {Flux density calibration};
    \node[action4, below = of fluxcal] (obsdata) {Obs data};
    \node[action4, below of =osg, node distance = 3.2cm] (tr) {Training data on CyVerse};
    \node[action, below of =apcalmemo, , node distance = 3.17cm] (nn) {\textsc{Zingularity} neural network};
    \node[input, below = of obsdata] (posteriors) {Posterior draws from bootstrapped data};
    
    \path [line] (grmhd) -- (grrt);
    \path [line] (grrt) -- (symba);
    \path [line, dashed] (obs) -- node[midway, above, sloped]{metadata} (vlbimon);
    \path [line, dashed] (vlbimon) -- node[midway, below, sloped, align=left]{atmospheric\\parameters} (meqsil);
    \path [line, dashed] (rpicardparams) -- (rpicard);
    \path [line, dashed] (rpicardparams) -- (rpicardsim);
    \path [line, dashed] (apcalmemo) -- (fluxcal);
    \path [line, dashed] (apcalmemo) -- node[midway, above, sloped, align=left]{bootstrap errors} (posteriors);
    \path [line, dashed] (apcalmemo) -- node[midway, above, sloped, align=left]{\hspace{2.7cm}antenna\\\hspace{2.7cm}parameters} (meqsil);
    \path [line] (obs) -- (corr);
    \path [line] (corr) -- (rpicard);
    \path [line] (rpicard) -- (fluxcal);
    \path [line] (fluxcal) -- (obsdata);
    \path [line] (obsdata) -- (posteriors);
    \path [line] (symba) -- (osg);
    \path [line] (osg) -- node[midway, right, align=left]{$\sim$$10^6$ datasets} (tr);
    \path [line] (tr) -- node[midway, above, sloped, align=left]{\texttt{TFRecords}} (nn);
    \path [line] (nn) -- (posteriors);
    
\end{tikzpicture}}
\caption{Flowchart of the \z{} data streams. The left column shows the pathway from the input theory models $\widetilde{M}$ to the training data $\widetilde{T}$. The right column shows the processing chain for the observational \ac{eht} data $\widetilde{U}$. The central column presents the common metadata used.}\label{fig:flowchart}
\end{figure*}
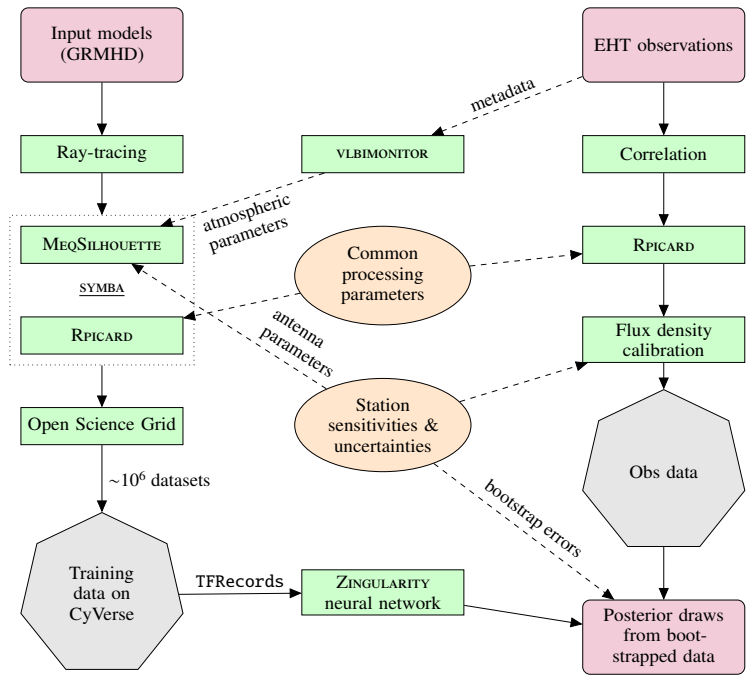

In this work, we used the EHT synthetic data library $\widetilde{S}$ that is based on the standard \ac{grmhd}-\ac{grrt} \sgra and \m87 models $\widetilde{M}$ from \citet{zingularity1} as training data $\widetilde{T}$ for \z{}.
A direct comparison of the models with EHT image reconstructions is not possible, because the observational images have a limited resolution and are not unique \citep{eht-paperIV, eht-SgrAiii}.

$\widetilde{S}$ was created with \symba \citep{2020Roelofs} to model the complete signal path of observational \ac{vlbi} data and stored as \texttt{TFRecord} files.
The $\widetilde{M}$ parameters of interest are the MAD/SANE magnetic state of the accretion disk $\phi_\mathrm{mag}$, the black hole spin $a_*$, the proportionality between the ion- to electron temperature $R_\mathrm{high}$ in the accretion disk, and for \sgra, also the inclination angle $i_\mathrm{los}$ and position angle $\theta_\mathrm{PA}$ of the source.
For \m87, we fixed $i_\mathrm{los} = \ang{17}$ \citep{2016Mertens} and $\theta_\mathrm{PA} = \ang{288}$ \citep{Walker2018}. We set black hole masses of $4.14 \times 10^6\,M_\odot$ and $6.2 \times 10^9\,M_\odot$ and distances of $8.127 \times 10^3$\,parsec (pc) and $16.9 \times 10^6$\,pc for \sgra, respectively \m87 \citep{2011Gebhardt, 2019Gravity, 2019Do}. Random variations at the 10\,\% level were added to the mass over distance ratios in the synthetic data generation as described in Section~3.6 of \citet{zingularity1}.
For each $\widetilde{M}$, we have about 1000 images. For \m87, we have multiple $\widetilde{S}$ realizations for each single image. For \sgra, we have multiple realizations from groups of 432 images, capturing the variability of the source during an observation.

For each realization, data corruption effects $\widetilde{C}$ due to thermal noise, uncertain telescope gains and polarization leakages, Earth's atmosphere, as well as the interstellar scattering screen toward \sgra were varied.
Given a set bandwidth, integration time, and quantization efficiency of the recorded data, the thermal noise was determined by the System Equivalent Flux Densities (SEFDs) of two antennas forming a baseline. The SEFD is the sum of all noise contributions along the signal path. Telescope gains are errors in the measured amplitude and phase of the signal. Polarization leakage is the cross-talk between the two orthogonal telescope receivers, which measure different polarization states of the incoming radiation. The Earth's atmosphere causes an attenuation of the signal, additional noise, and phase errors. The \sgra scattering screen leads to a blurred source image with additional induced substructures. These corruption effects are described in detail in Section~4 of \citet{zingularity1}.

As noted in \Cref{sec:intro}, some regions of the \ac{grmhd}-\ac{grrt} parameter space are disfavored based on simple comparisons with the \ac{eht} data and multiwavelength constraints. Nonetheless, we did not apply any a priori cuts on $\widetilde{T}$ and trained on data that covers the full $\widetilde{M}$ parameter space. On the one hand, this serves as a validation for \z{}, as the posterior probability from the fit should be disjoint from the the parameter space that is strongly disfavored by \ac{eht} constraints applied in earlier works.
On the other hand, our strategy serves as a test of the models.
If \ac{grmhd}-\ac{grrt} describes the physical reality of \sgra and \m87 well, the models favored by \z{} fits should be in agreement with (quasi-)simultaneous multiwavelength constraints in the absence of parameter degeneracies.

\Cref{tab:zingularity} gives an overview of the training data and neural network parameters used in this work. We have $N_\mathrm{tr}$ individual $\widetilde{T}$ samples. Each consists of $N_\mathrm{vis}$ real and imaginary values of the complex correlation coefficients per RR, RL, LR, and LL correlation product.
The $(u, v)$ coverage corresponds to the 2017 \ac{eht} observations on April 7 for \sgra and April 11 for \m87 \citep{eht-paperII, zingularity1}.
The time cadence of the data within \ac{vlbi} scans is kept to a short 10\,s sampling, to avoid decoherence effects from residual phase errors.
Using the correlation outputs directly without long averaging intervals keeps the systematic error budgets low; the determined total intensity error budget can be erroneous in the presence of circularly polarized source signals and long averaging times likely cause decoherence in mm VLBI observations. For the current \ac{eht} data of \sgra in particular, both effects need to be taken into account \citep{eht-SgrAii}. But also for the \m87 \ac{eht} data, coherence losses will occur when averaging the data in time \citep[Figure 18 in][]{eht-paperIII}.

\Cref{fig:training_data} shows a few $\widetilde{T}$ examples alongside the underlying ground-truth $\widetilde{M}$.
Unlike the native visibilities used as neural network input, the data shown here are in a physically meaningful Stokes $\mathcal{I}$ (total intensity), $\mathcal{Q} \& \mathcal{U}$ (linear polarization), $\mathcal{V}$ (circular polarization) amplitude and phase representation averaged over VLBI scan durations of a few-minutes.
The \sgra data have a better $(u, v)$ coverage and exhibit intrinsic variability -- while a single static frame made up an \m87 dataset, multiple frames were used akin to a movie for \sgra, as the hours-long EHT observing track is much longer than the $\sim$\,20\,s gravitational timescale of the source. 

The two \m87 models displayed differ in the black hole spin parameter $a_*$. The model with the highly spinning $a_*=0.94$ black hole possesses more extended emission. With relatively low $R_\mathrm{high} = 10$ values, the ratio of jet to disk emission is comparatively small. 
Between about 3.5\,G$\lambda$ and 4.5\,G$\lambda$, a clusters of Stokes $\mathcal{I}$ visibility phases are offset by roughly \ang{180} between the two models. Concurrently, slight offsets and a steeper phase evolution with baseline length for the $a_*=0.5$ model data are present for the $\mathcal{Q}$ and $\mathcal{U}$ phases. The collective differences in phases across multiple Stokes parameters can be possible salient model features that allow for a distinction of the spin parameter in this case study. Other differences in the visibility data could be the result of different $\widetilde{C}$ realizations. For example, telescope gain errors could cause the differences of Stokes $\mathcal{I}$ amplitudes, while polarization leakage could be responsible for the $\mathcal{Q}$ and $\mathcal{U}$ phase differences at other baseline locations, where Stokes $\mathcal{I}$ shows no significant changes.

Out of the few \sgra $\widetilde{M}$ that pass most of the multiwavelength and EHT data constraints considered in \citet{eht-SgrAv}, we selected two examples for \Cref{fig:training_data}. The SANE $a_*=0.5$, $R_\mathrm{high} = 40$, $i_\mathrm{los} = \ang{10}$ model fails only the 86\,GHz source size. The MAD $a_*=0.94$, $R_\mathrm{high} = 160$, $i_\mathrm{los} = \ang{30}$ model fails only the variability constraints \citep{eht-SgrA_LC, 2022Broderick}.

For the \sgra data, it is more difficult to identify salient features by eye due to the intrinsic $\widetilde{M}$ variability.
A possibly important feature is the higher polarization of the MAD model; across all baseline lengths, the $\mathcal{Q}$ and $\mathcal{U}$ amplitudes are higher and phases more coherent. As noted in \citet{eht-SgrAv}, the \sgra SANE model shown here is indeed most likely weakly polarized. However, the polarization can differ between model frames and be affected by polarization leakage, which makes the need for a deep BANN, trained on many $\widetilde{M}$ and $\widetilde{C}$ realizations to identify the robust salient features, evident.

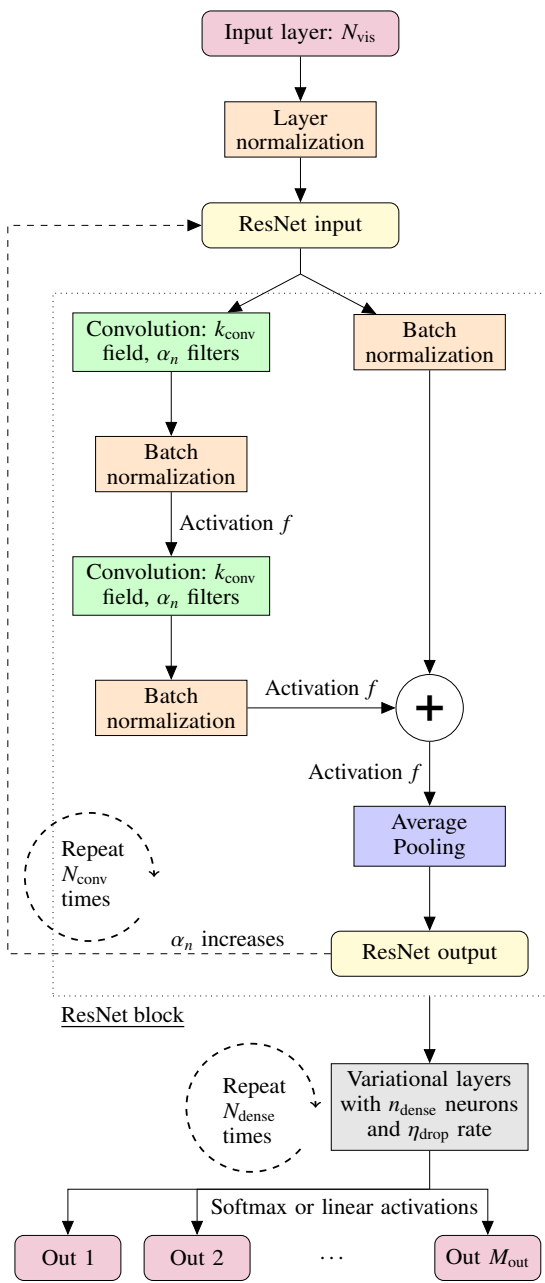
\begin{figure}
\centering
\scalebox{.85}{
\tikzset{
    between/.style args={#1 and #2}{
         at = ($(#1)!0.5!(#2)$)
    }
}

\tikzstyle{action} = [draw, rectangle,fill=green!20, node distance=1cm,
    minimum height=2em, text width=8em, text centered]
\tikzstyle{action2} = [draw, rectangle,fill=orange!20, node distance=1cm,
    minimum height=2em, text width=6em, text centered]
\tikzstyle{action3} = [draw, rectangle,fill=blue!20, node distance=1cm,
    minimum height=2em, text width=6em, text centered]
\tikzstyle{action4} = [draw, rectangle,fill=gray!20, node distance=1.35cm,
    minimum height=2em, text width=8em, text centered]
\tikzstyle{input} = [rectangle, draw, fill=purple!20, 
    text width=8em, text centered, node distance=0.5cm and 7.5cm, minimum height =2em, rounded corners]
\tikzstyle{data} = [rectangle, draw, fill=yellow!20, 
    text width=8em, text centered, node distance=1cm and 7.5cm, minimum height =2em, rounded corners]
\tikzstyle{output} = [rectangle, draw, fill=purple!20, 
    text width=4em, text centered, node distance=1cm and 7.5cm, minimum height =2em, rounded corners]
\tikzstyle{addcircle} = [circle, minimum size=0.6cm, text centered, draw=black, fill=white, text width=0.6cm, align=center]
\tikzstyle{line} = [draw, -triangle 45]
\tikzstyle{line2} = [draw]

\begin{tikzpicture}[node distance = 2cm, auto]
    \node [input] (input) {Input layer: $N_\mathrm{vis}$};
    \node [action2, below = of input, yshift=0.3cm] (lnorm) {Layer normalization};
    \draw[thick, dashed, ->] (-2.5,-13.8) arc (320:0:1);
    \node[text width=3cm] at (-2.2,-13.1) 
    {Repeat\\$N_\mathrm{conv}$\\times};
    \node [data, below = of lnorm, yshift=0.3cm] (rinp) {ResNet input};
    \coordinate[below=4mm of rinp.south] (aux1);
    \node [action, below = of rinp, xshift=-2cm] (conv1) {Convolution: $k_\mathrm{conv}$ field, $\alpha_n$ filters};
    \node [action2, below = of conv1] (norm1) {Batch normalization};
    \node [action, below = of norm1] (conv2) {Convolution: $k_\mathrm{conv}$ field, $\alpha_n$ filters};
    \node [action2, below = of conv2] (norm2) {Batch normalization};
    \node [action2, right of=conv1, xshift=3cm] (skip_norm) {Batch normalization};
    \node (add) [addcircle, right of=norm2, xshift=2cm] {\huge\textbf{+}};
    \node[action3, below = of add] (avg) {Average Pooling};
    \node[data, below = of avg] (rout) {ResNet output};
    \node[draw,dotted, inner sep=3mm,label={[xshift=-2.75cm, yshift=-11.6cm]\underline{ResNet block}},fit=(conv1) (rout) (skip_norm)] (resnet) {};
    \node[text width=3cm] at (-0.5,-14.15) 
    {$\alpha_n$ increases};
    \node [action4, below = of rout] (ds) {Variational layers with $n_\mathrm{dense}$ neurons and $\eta_\mathrm{drop}$ rate};
    \draw[thick, dashed, ->] (0,-17.4) arc (320:0:1);
    \node[text width=3cm] at (0.3,-16.8) 
    {Repeat\\$N_\mathrm{dense}$\\times};
    \coordinate[below=0.6cm of ds.south,] (aux2);
    \node [output] at (-3.6,-19.1) (out1) {Out 1};
    \node [output] at (-1.6,-19.1) (out2) {Out 2};
    \node [output] at (2.9,-19.1) (outM) {Out $M_\mathrm{out}$};
    \node[text width=2cm] at (1.3, -19.1) 
    {\centering{\ldots}};
    \node[text width=5cm] at (1.11,-18.3) 
    {Softmax or linear activations};

    \path [line] (input) -- (lnorm);
    \path [line] (lnorm) -- (rinp);
    \path [line2] (rinp.south) -- (aux1);
    \path [line] (aux1.south) -- (skip_norm);
    \path [line] (aux1.south) -- (conv1);
    \path [line] (conv1) -- (norm1);
    \path [line] (norm1) -- node[midway, right, align=left]{Activation $f$} (conv2);
    \path [line] (conv2) -- (norm2);
    \path [line] (norm2) -- node[midway, above, align=left]{Activation $f$} (add);
    \path [line] (skip_norm) -- (add);
    \path [line, dashed] (rout.west) -- ++(-5,0) |- (rinp.west);
    \path [line] (add) -- node[midway, left, align=left]{Activation $f$} (avg);
    \path [line] (avg) -- (rout);
    \path [line] ([xshift=2cm]resnet.south) -- (ds.north);
    \path [line2] (ds.south) -- (aux2);
    \path [line] (aux2) -- ++(0,0) -| (out1.north);
    \path [line] (aux2) -- ++(0,0) -| (out2);
    \path [line] (aux2) -- ++(0,0) -| (outM);
    
\end{tikzpicture}}
\caption{Layout and data flow of our chosen \m87 and \sgra BANN architectures. The ResNet and dense variational blocks are repeated $N_\mathrm{conv}$ and $N_\mathrm{dense}$ times, respectively. Single variational neurons are used in each output layer.}\label{fig:architecture}
\end{figure}

\section{Zingularity}
\label{sec:zingularity}

\z\footnote{\url{https://gitlab.com/mjanssen2308/zingularity}} is a modular open-source framework for the implementation of Bayesian \tensorflow{}-based neural networks.
The input data for $\widetilde{T}$ and $\widetilde{U}$ is converted into the optimized and efficient \texttt{TFRecord} format (see Section~5 in \citet{zingularity1}).
Based on this self-contained uniform data format, any kind of \tensorflow{} ANN can be trained and used for inference, independent from the type of input data.
\z{} runs on CPUs, a single GPU/TPU, or through distributed computing on multiple GPUs/TPUs.
Bootstrapping methods, where $\widetilde{U}$ is resampled based on known $\widetilde{C}$, are implemented as well. For each resampled dataset, the surrogate posterior $q_\varphi$ can be computed efficiently for inference.

\Cref{fig:flowchart} shows an overview of the complete \z{} processing chain from the input data to posteriors; how the training data are produced from the GRMHD models, how the observational data are processed, and the parallels between the observation and theory processing chains. The integration of the $\textsc{vlbimonitor}$ \citep{eht-paperII}, Open Science Grid \citep{osg07, osg09}, and CyVerse data storage \citep{cyverse1, cyverse2} into a single pipeline with the \textsc{Pegasus} software \citep{deelman-fgcs-2015} is described in \citet{zingularity1}.

\subsection{EHT GRMHD-GRRT network implementation}
\label{sec:eht_grmhd_grrt_network_implement}

In this work, we used a BANN architecture that couples a residual (ResNet) Convolutional Neural Network \citep{cnn1, cnn2, ResNet} with several \texttt{DenseVariational} fully connected layers performing variational inference (\Cref{fig:architecture}).
The implementation  in \z{} is done with \tensorflowprobability{} \citep[][\Cref{sec:hidden-layers}]{tf-probability}, combined with multiple regularization methods for an increased model robustness (\Cref{sec:overfitting}).

A deep architecture makes it possible for the network to pick up complex data combinations as salient features, closure quantities for example.
With the use of variational layers with trainable weight distributions as hidden- and output layers, the BANN captures the epistemic and aleatoric uncertainties. These correspond to uncertainties in the model and the data, respectively.

In the following subsections, we describe our BANNs used to infer GRMHD-GRRT parameters from EHT observations as implemented in \z{}. The numerical values and method implementations for all relevant parameters of the network are listed in \Cref{tab:zingularity}.

\subsubsection{Input and output}

\z{} performs the conversion of the input data to \texttt{TFRecord} files as a pre-processing step. The flexible conversion pipeline can handle any type of labeled or unlabled data.
For full-polarization interferometric data, the real and imaginary component arrays for each correlation product correspond to 8 input `channels' in our neural network architecture.
An initial layer normalization \citep{LayerNormalization} is applied to the input, where the normalization parameters are optimized using each single example in a training batch.

The output layers consist of fully connected single Bayesian variational neurons for each inference task: With linear activation functions for $a_*$, $R_\mathrm{high}$, and for \sgra also $i_\mathrm{los}$ as well as $\theta_\mathrm{PA}$.
The $\phi_\mathrm{mag}$ classification is done with a softmax activation. The priors and posterior functions used for the regression output layers are the same as those used by the hidden layers. The hidden layers (\Cref{sec:hidden-layers}) model the data between the input and output.
We use a single chain of hidden layers and split only in the last output layer. All of our $M_\mathrm{out}$ output layers are connected to same last hidden layer, so we can infer dependences between $\widetilde{M}$ parameters. We have tested that using individual, distinct networks per parameter does not improve the inference accuracy.

When applied to \sgra and \m87, the BANN is trained for $N_\mathrm{ep}$ epochs (iterations over $\widetilde{T}$). $N_\mathrm{ep}$ is determined empirically as the value where the loss $L$ saturates. As studied by \citet{Nep}, the best stopping criterion is where the loss curve starts developing a very shallow, plateau-like curve. When training for longer, the network may fit to noise in the training data.
In practice, we determine the approximate saturation point by eye for each model and then survey a range of $N_\mathrm{ep}$ around it. For \sgra we find two equally viable $N_\mathrm{ep}$ of 50 and 60. Unless stated otherwise, we are using the $N_\mathrm{ep} = 60$ trained network as our fiducial \sgra model for the analyses presented in this work.

The hidden layers between the input and output are described below.

\subsubsection{Hidden layers}
\label{sec:hidden-layers}

Generally, we need a BANN with sufficiently large trainable parameters $N_\mathrm{free}$, to be in an overparameterized ``double descent'' regime \citep[e.g.,][]{dd1, dd2}. Additionally, we strive for deeper rather than wider networks by having more layers and fewer neurons per layer.
In practice, our network widths are based on the input data size and computational limitations. Network depths are increased until no further improvements in validation errors are gained.
We experimented with different BANN architectures and selected the one with the best performance in our parameter surveys described later in this manuscript.

The combination of convolution operations and skip connections in the ResNet enables an in-depth modeling of the data to pick out the salient features in the data (i.e., which locations in time-baseline space and combinations of the visibility data are the most informative for the $\widetilde{M}$ parameter inference in the presence of $\widetilde{C}$).
The Bayesian nature of the variational layers enables us to see parameter dependences and uncertainties in the posteriors formed from the ResNet pre-processed data.

\paragraph{ResNet blocks:}

The input visibilities are time-baseline sorted into a two-dimensional $(8, N_\mathrm{vis})$ shape. Initially, the data flows through $N_\mathrm{conv} = \log_2{N_\mathrm{vis}}$ ResNet blocks.

Each block has the same architecture, where the data are passed through two parallel branches.
The first branch consists of two consecutive and identical sub-blocks. In these sub-blocks, the data first flows through a convolution layer with $\alpha_n$ filters/neurons, then a Batch normalization layer (which normalizes the input per training batch), and finally a layer with activation function $f$. For the ResNet block number $n = 1, 2, ..., N_\mathrm{conv}$, we have
\begin{equation}
    \alpha_n = \left(n_\mathrm{CNNb}\right)^{1-\frac{n}{N_\mathrm{conv}}} \left(n_\mathrm{CNNl}\right)^{\frac{n}{N_\mathrm{conv}}} \,. 
\end{equation}
Each convolution filter $\alpha_n$ has a receptive field of size $k_\mathrm{conv}$.
The purpose of these operations is to extract the most meaningful locations and combinations of the data in terms of information content. We thus refer to this as modeling branch.

The second branch is a skip connection, where the data flows through one layer with $\alpha_n$ convolution filters, each having a receptive field of one (to have a matching dimension of the data with the modeling branch), followed by Batch normalization.

The output of the two branches are then added and passed through another activation $f$.
Finally, the data are downsampled by a factor of two in an average pooling layer.
The data in the last ResNet block will have a dimension equal to $n_\mathrm{CNNl}$.

The initial weight parameters of each convolution layer are drawn from a uniform distribution between $\pm \sqrt{6 / N_\mathrm{I+O}}$, where $N_\mathrm{I+O}$ is the number of input plus output units \citep{2010Glorot}. Bias terms are initialized as zeros.

\paragraph{Bayesian fully connected layers:}

After the ResNet blocks, the data are passed through $N_\mathrm{dense}$ fully connected variational layers, each having $n_\mathrm{dense}$ neurons with activation function $f$.

Multivariate standard Normal distributions serve as prior for the weights, while no variational inference is performed for the bias terms. The surrogate posteriors are represented with trainable Normal distributions.

We also tried network architectures that consisted only out of Bayesian fully connected layers, without any convolution layers. Overall, these had higher validation errors compared to our fiducial ResNet+fully connected models. Yet, for \m87 the parameter inference on observational data gave consistent results \citep{zingularity3}. For \sgra, it was difficult to find robust network architectures without convolutions through our parameter survey method (\Cref{sec:zingularitysurvey}).
\vspace{0.1cm}
\subsubsection{Generalization}
\label{sec:overfitting}

Overfitting is common in ML applications, as a model may train itself on peculiarities of $\widetilde{T}$, which are not generally representative for the underlying features of interest from $\widetilde{M}$.
Overfitting in the traditional statistical sense can however be benign in deep learning tasks: Neural networks can perform well, even with perfect fits to training data when the sample size is much smaller than the number of possible directions in parameter space that are unimportant for predictions \citep[e.g.,][]{benign_overfit1, benign_overfit2}. We strive for a model that generalizes well.

We split out a fraction $\eta_\mathrm{val}$ of $\widetilde{T}$ as validation data. These data are not used for training and can therefore be seen as uncharacterized data $\widetilde{U}$ that is unknown to the network. We check that our performance metrics (\Cref{sec:metrics}) give the same answers for the training and validation data.

Additionally, a Dropout regularization is employed, where a fraction of $\eta_\mathrm{drop}$ of each hidden layers' neurons are randomly deactivated in each forward pass when the network is trained and inferences are made \citep{2012arXiv1207.0580H, pmlr-v28-wan13}.
The stochastic drops of neurons are akin to a random sampling over different network architectures, thereby reducing noise in the trained parameters.

Finally, $L_1$ and $L_2$ regularization losses are added to $L$ with rate hyperparameters of $\mathcal{L}_1$ and $\mathcal{L}_2$, respectively.
By penalizing the absolute value of the weights in the loss function, $L_1$ steers weights of neurons toward zero. As a result, the sparsity of the network increases, as unimportant data pathways can be deactivated.
$L_2$ penalizes the sum of the weights' squares, steering them toward smaller values. Smaller weights reduce the complexity of the network and single weight outliers that have high values will be strongly reduced.

In this work, we are surveying only small values of $\eta_\mathrm{drop}, L_1,$ and $L_2$. Larger values typically cause larger training losses without improving validation errors.
\vspace{0.1cm}
\subsubsection{Optimization algorithm}

We use the \texttt{RMSProp} algorithm for the gradient descent along the loss surface of our deep BANN, which is unlikely to get stuck in a bad local minima \citep{2014Choromanska, 2016Kawaguchi}.
With the simple stochastic gradient descent, we encountered problematic overfitting, where the training/validation loss was decreasing/increasing.
\texttt{RMSProp} keeps a moving average of the square gradients and divides the gradient by the root of this average. This enables a dynamic learning rate along each dimension of the gradient descent, which is not significantly slowed down by gradients from past iterations.
To increase the stability of the network training, we also use an adaptive overall learning rate with a peak value of $l_r$.
Following \citet{2016Loshchilov}, \citet{2017Goyal}, and \citet{2018He}, we use an initial linear learning rate warm-up from $10 l_r / N_\mathrm{ep}$ at the first epoch to $l_r$ when 10\,\% of the epochs have been processed, followed by a cosine decay.
The rationale is to have an initially low learning rate for numerical stability of the training of a network with initially random parameters and to smoothly decrease the learning rate when the BANN parameters are converging.
The gradient descent iteration through $\widetilde{T}$ is done in small batches of size $N_\mathrm{b}$.
\vspace{0.1cm}
\subsubsection{Performance metrics}
\label{sec:metrics}

We use a mean absolute error metric to track the regression performance of the network during the training for both the training and validation data.
For the classification, we measure the fraction of correct predictions over the total number of predictions as the network's accuracy.

We typically report only the average performance metrics for all training and validation $\widetilde{T}$. In principle, it would also be possible to look at the network performance for distinct $\widetilde{M}$ parameter regions. Yet, parameter-dependent network performances will be evident in the $\widetilde{U}$ posteriors, which are the final information of interest; if the inferred parameters are in a region of good/poor network performance, the posteriors will be narrow/wide.
In the same vein, the BANN can deal with parameter degeneracies.
\vspace{0.1cm}
\subsection{Bootstrapping of known data corruption effects}
 
Within the generic \z{} bootstrapping framework, we resample uncharacterized EHT visibilities with $\widetilde{C}$ in the following order:
\begin{enumerate}
    \item Per-antenna polarization leakage terms $\mathcal{D}$ \citep{eht-m87-paper-vii}.
    \item Per-antenna amplitude gain errors. Typically, these are assumed to be $\mathcal{O}(10\,\%)$ for all stations. Here, we adopt a more precise description. From the gain (DPFU) measurements of each antenna, we take the determined statistical $\mathcal{G}_\mathrm{scatter}$ uncertainty from the measurement error into account. Separately, we group the APEX, IRAM\,30m, LMT, and SPT stations and add a common $\mathcal{G}_\mathrm{planet}$ uncertainty based on the accuracy of the model used for the common solar system object (Saturn) that served as a primary calibrator.
    For all other stations, $\mathcal{G}_\mathrm{planet}$ is added independently as different primary calibrators were used \citep{Janssen2019b}.
    \item Per-antenna gain curve uncertainties. For the fitted gain curves as a function of elevation $E$, $\mathrm{gc}(B, E_0;\, E) = 1 - B \left( E - E_0 \right)^2$, statistical $\mathrm{gc}_B$ and $\mathrm{gc}_{E0}$ uncertainties from the fit are taken into account \citep{Janssen2019b}. These are usually ignored, but they can become important for data taken at low elevations.
    \item Per-baseline thermal noise. These are taken from the CASA \texttt{SIGMA} estimator of the visibilities themselves \citep[see, e.g., Section 2 of][]{Janssen2019}.
\end{enumerate}

The magnitude of these corruptions is given in \Cref{tab:zingularity}.
For each resampled dataset, we draw many times from the network's posterior (a forward pass takes only a few seconds of computational time) to take the combined uncertainties into account and obtain conservative results. We find bootstrapping to be useful for ensuring that the trained BANN does not overfit on data corruption effects.
\vspace{0.1cm}
\subsection{Software architecture and scientific reproducibility}

We host \z{} on \textsc{GitLab}.\footnote{Accessible under \\ \url{https://gitlab.com/mjanssen2308/zingularity}}
The software is freely available under the GNU General Public License. All configuration options are set with a single YAML input file. We implemented a deterministic random number seeding for \tensorflow{} in \z{}.

Using a Continuous Integration / Continuous Delivery setup, with every code change we automatically run unit tests, deploy a containerized \textsc{Docker}\footnote{\url{https://www.docker.com/}} version of the software, and generate documentation with \textsc{Doxygen}.\footnote{\url{https://www.doxygen.nl/}}
A \textsc{Docker} container for every git commit hash is available online.\footnote{\url{https://hub.docker.com/r/mjanssen2308/zingularity}}

The results shown here are derived with \z{} version 1.0.0, which is based on \tensorflow{} version 2.5.0.
The \textsc{Docker} container \texttt{36d816a5d063e673f7502a8aa2eaf4a870431a02}\footnote{\url{https://tinyurl.com/3m49tm7e}} can be used to reproduce the results with the \texttt{zingularity-EHT2017} configuration files for \sgra and \m87 that are stored in the container under \texttt{/usr/local/src/zingularity/input\_examples}.
The synthetic data used for the network training in this work are described in \citet{zingularity1}.

\begin{figure*}
    \centering
    \begin{tabular*}{\textwidth}{@{}l@{\hspace{0.01\textwidth}}c@{\hspace{0.01\textwidth}}r@{}}
    \includegraphics[width=0.325\textwidth]{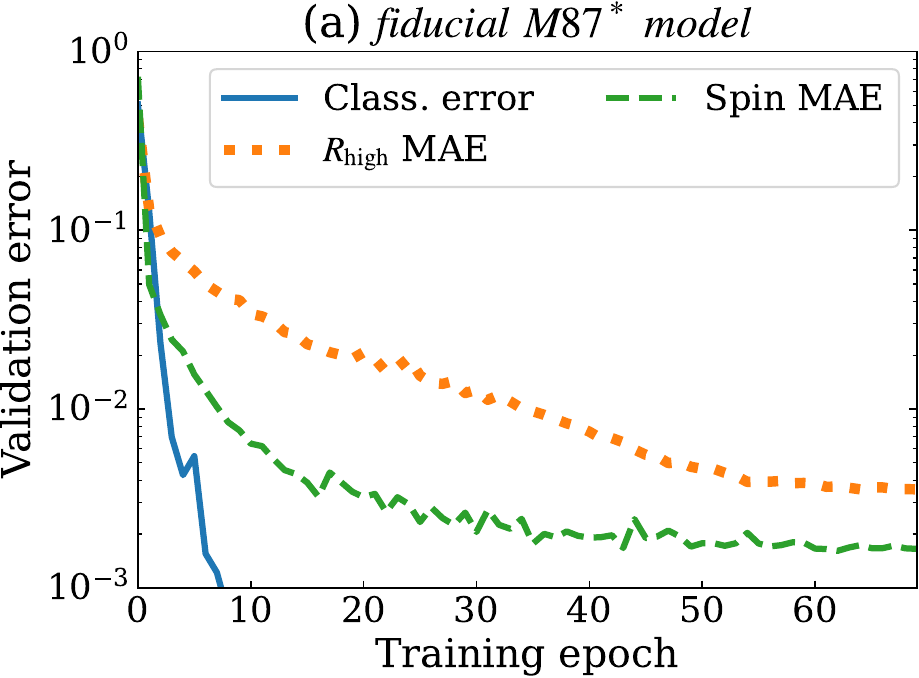} &
    \includegraphics[width=0.325\textwidth]{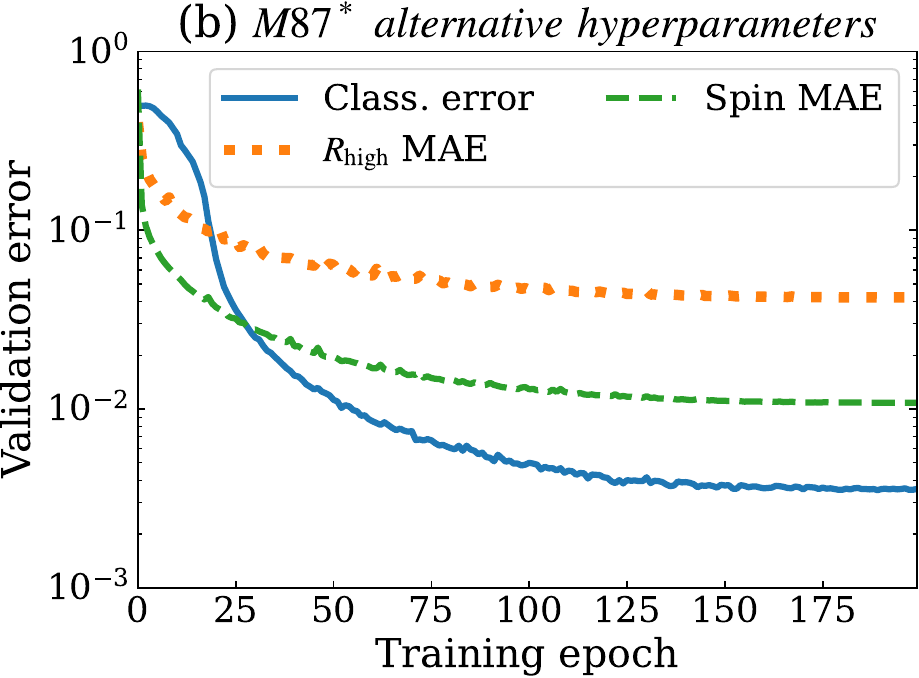} &
    \includegraphics[width=0.325\textwidth]{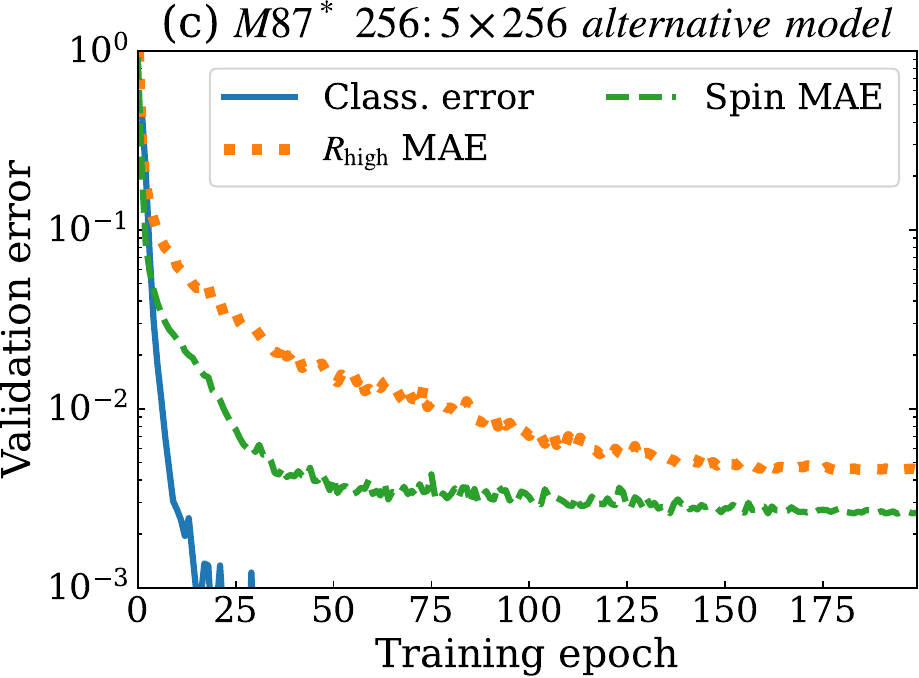} \\
    \includegraphics[width=0.325\textwidth]{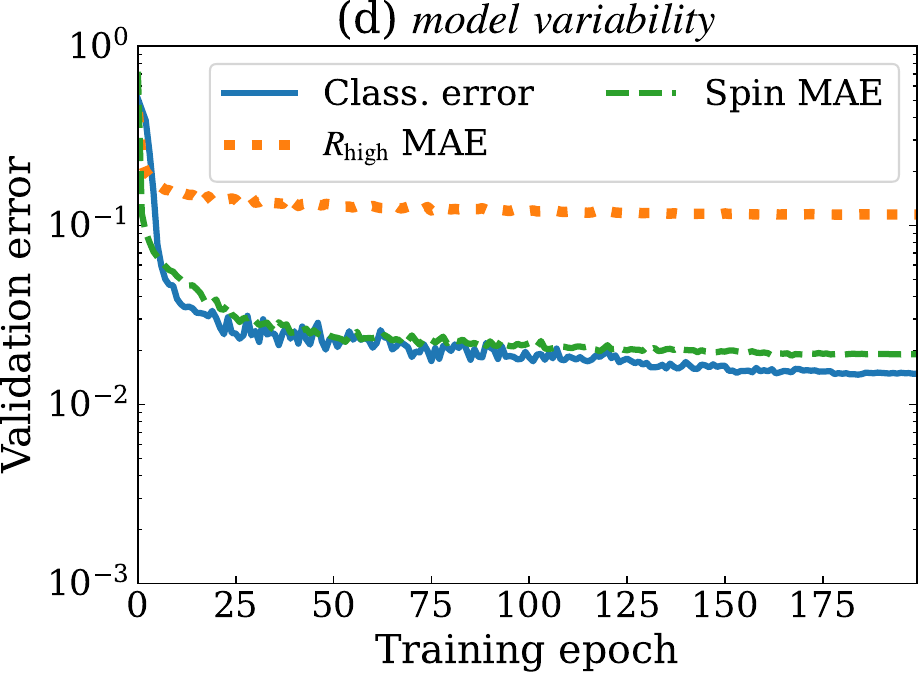} &
    \includegraphics[width=0.325\textwidth]{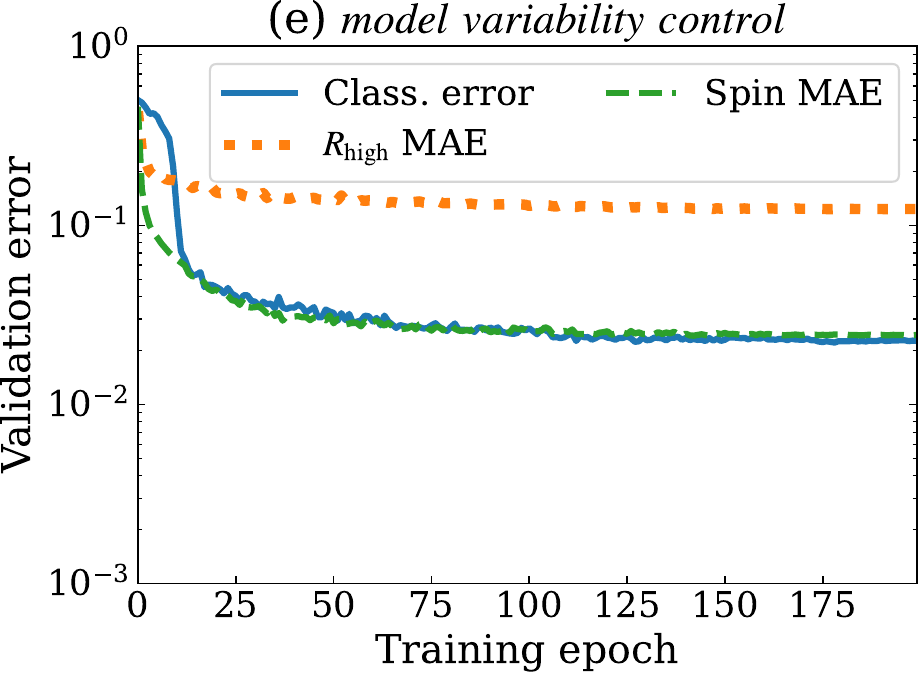} &
    \includegraphics[width=0.325\textwidth]{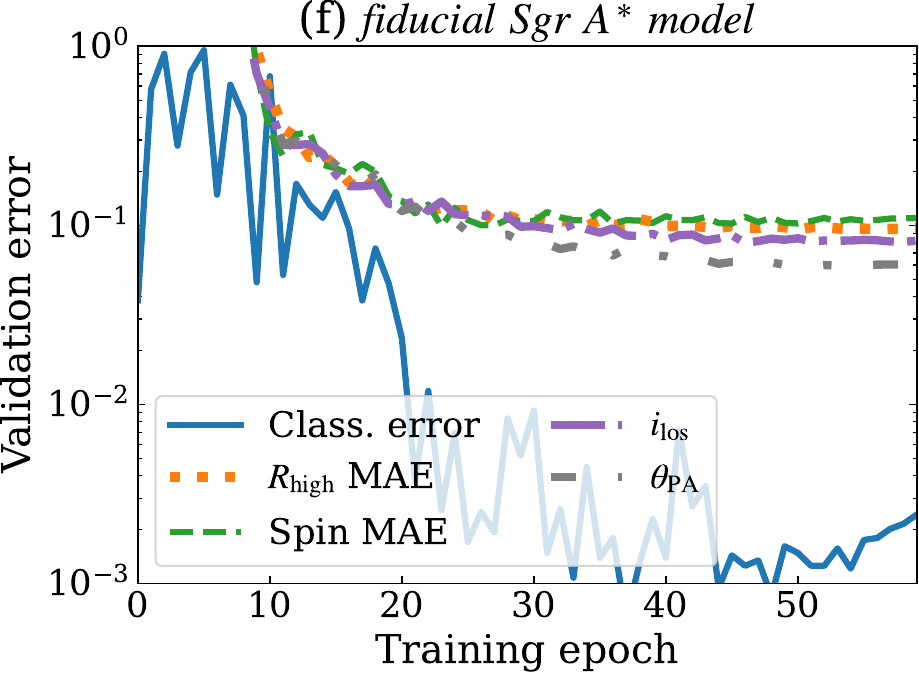} \\
    \includegraphics[width=0.325\textwidth]{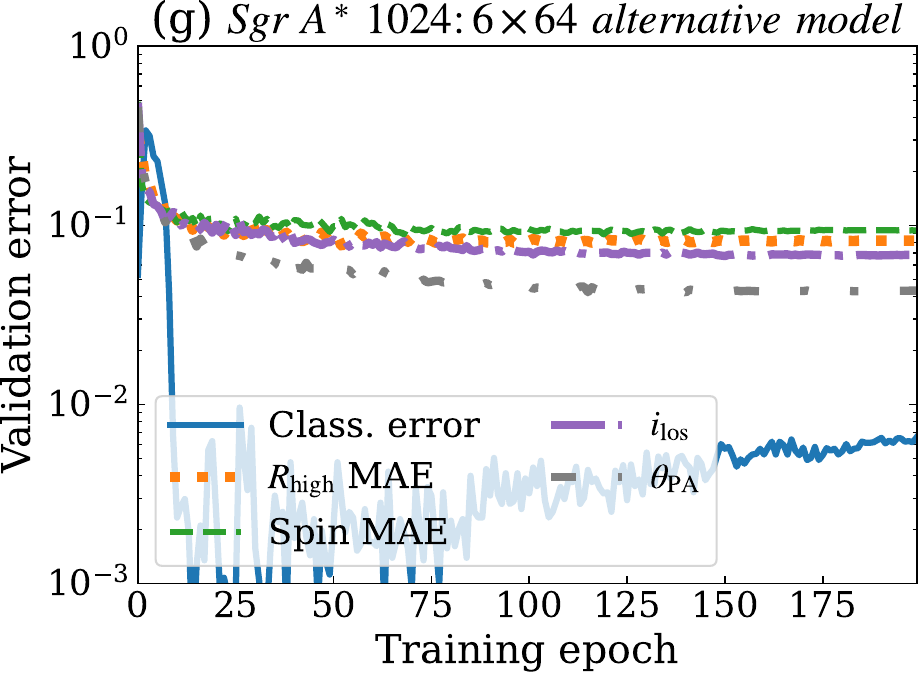} &
    \includegraphics[width=0.325\textwidth]{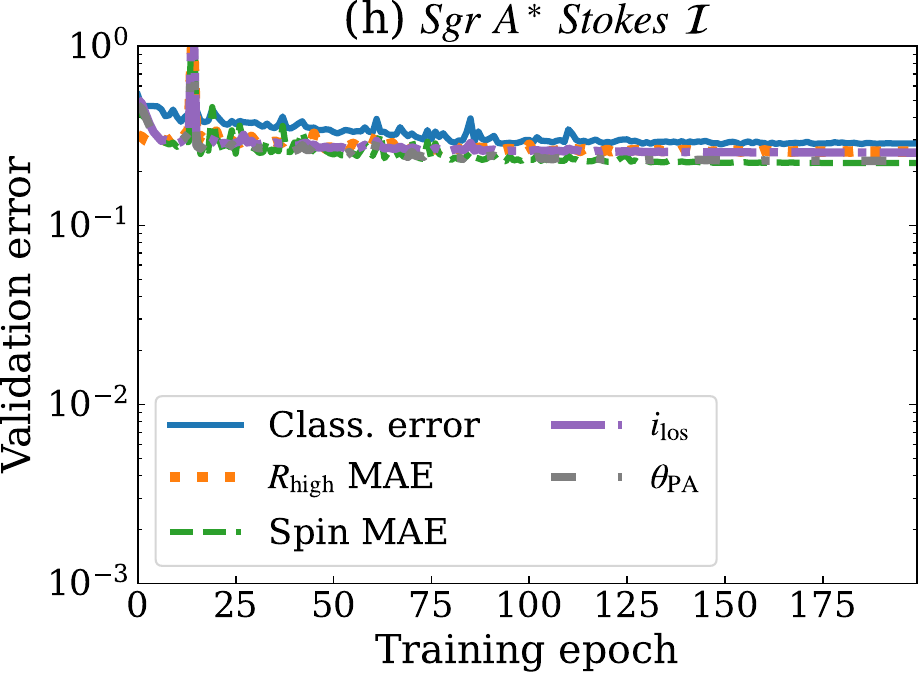} &
    \includegraphics[width=0.325\textwidth]{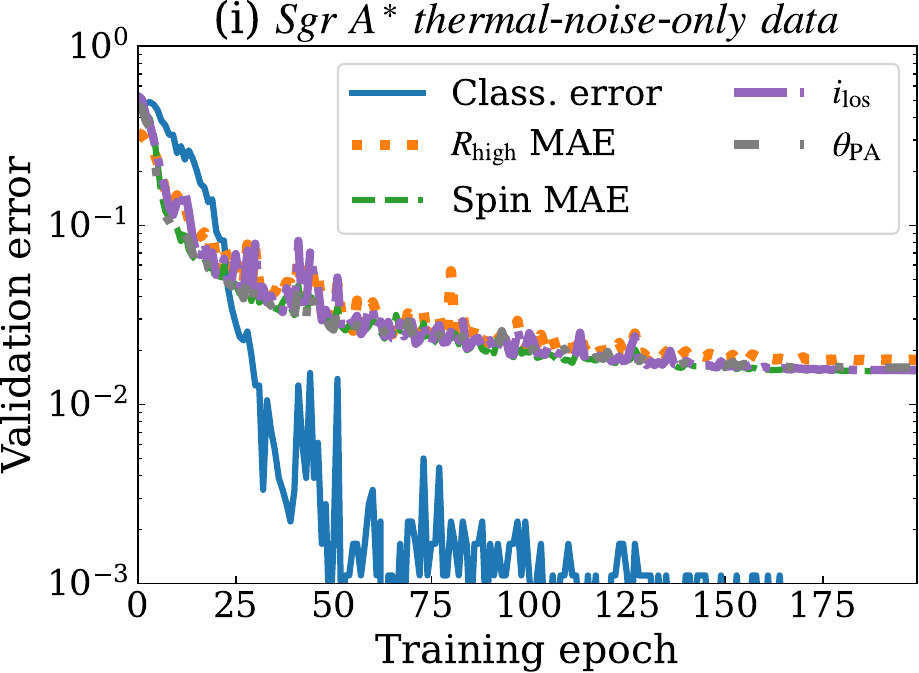}
    \end{tabular*}
    \caption{Performance metrics for the \sgra and \m87 network training are displayed for various dedicated \z{} validation tests as described in \Cref{sec:zingularity-validation}. The validation error is computed from normalized labels of validation data not seen by the network during training. The mean absolute error (MAE) is computed as the average over all validation samples for normalized regression labels. The classification error (Class. error) is defined as one minus the network's accuracy, i.e., the fraction of misclassified validation samples. For some studies, the classification errors get numerically close to zero beyond the logarithmic y-axis limit displayed in the figure. Training errors (not shown here) follow the validation curves with a constant (negative) offset, as is typical for neural networks \citep[e.g.,][]{2017Advani}. For the \m87 data, $i_\mathrm{los}$ and $\theta_\mathrm{PA}$ are fixed. The training of the fiducial models is shown only up to the determined $N_\mathrm{ep}$.
    }
    \label{fig:zing_validation}
\end{figure*}
\vspace{0.1cm}
\subsection{Computational efficiency}
\label{sec:benchmark-tests}

The \tensorflow{} backend of \z{} is optimized for CPUs, GPUs, and TPUs.
Through the \tensorflow{} reader of \texttt{TFRecord} files, \z{} efficiently shuffles the input randomly, caches and pre-fetches data for multiple training epochs, and splits the features into small batches for the training. These optimizations result in an efficient interplay of GPU/TPU and CPU computations.

Moreover, we shard the \texttt{TFRecord} input, which enables parallel read access for the training.
This is utilized with \z{} through an \textsc{Horovod} \citep{horovod} implementation of distributed computing.
The \textsc{Horovod} parallelization speedup can be configured with the NVIDIA Collective Communication Library version 2 (NCCL 2) or any proprietary MPI implementation for GPUs that support the \texttt{allreduce} or \texttt{allgather}, \texttt{broadcast}, and \texttt{reducescatter} operations. The open-source MPI implementations are usually not as fast and, by default, are used only when NCCL is unavailable.
The \textsc{Horovod} Tensor Fusion offers an additional computational acceleration by batching together as many tensors, that are queued to be processed, as possible into a single \texttt{allreduce} operation.

Distributed computing benchmark tests presented in \citet{horovod} demonstrated the efficient scalability of \textsc{Horovod} over native \tensorflow{} on up to 128 GPUs. As a future outlook, we note the ongoing quantum machine learning developments of our underlying \tensorflow{} framework following the latest quantum hardware advancements \citep{TFquantum}.

GPU-based correlators offer fast and energy-efficient processing platforms for astronomical interferometers, which scale well for a large number of antennas. Novel correlator designs make use of the tensor core technology of new GPUs, which is yet more efficient \citep{2021Romein,galaxies11010013}. 
It is worth noting that such computing platforms would be perfectly suitable for \tensorflow{}-based applications such as \z{} during telescope downtime.

Taking one of our BANNs with 12 million trainable parameters as a representative example, we found that a full iteration (training epoch) over 600,000 datasets, each with 21,956 visibility data points, takes 30 seconds on a single NVIDIA A100 GPU. Obtaining 100 posterior samples from 100 bootstrapped $\widetilde{U}$ takes 20 seconds.
We have used the containerized version of \z{} through Apptainer \citep[formerly Singularity,][]{singularity1} for these tests.

\section{Validation}
\label{sec:zingularity-validation}

\subsection{Training diagnostics}

\z{} launches the \textsc{TensorBoard} visualization and diagnostics toolkit from \tensorflow{}. The dashboard shows the evolution of losses, performance metrics, as well as biases and weights for each layer. Additionally, a graph of the neural network can be displayed. Together with integrated features such as the \textsc{What-If-Tool} \citep{2019WIT}, \textsc{TensorBoard} is an Explainable AI feature, which helps with the identification of salient features in the data and understanding of the choices made by neural networks.
We have performed several targeted validation tests with the help of \textsc{TensorBoard} data. Unless stated otherwise, we have used the full standard training sets with 600,000 samples for \m87 and 252,000 samples for \sgra.
\Cref{fig:zing_validation} shows the results of these \z{} tests:
\begin{enumerate}[label=(\alph*), topsep=2ex, itemsep=1ex,partopsep=1ex,parsep=1ex]
    \item fiducial \m87 model -- we show the training performance metrics for our \ac{eht} 2017 \m87 model described in \Cref{sec:eht_grmhd_grrt_network_implement}.
    \item \m87 alternative hyperparameters -- we show how well our \ac{eht} 2017 \m87 model performs with a different set of $f = $ ReLU, $\Xi = $ stochastic gradient descent (SGD), $\eta_\mathrm{drop} = 0, \mathcal{L}_1 = 0, \mathcal{L}_2 = 0.05$ hyperparameters as opposed to those listed in \Cref{tab:zingularity}, using the same $n_\mathrm{CNNl}\!:N_\mathrm{dense} \times n_\mathrm{dense}$ layout. 
    \item \m87 $256\!:5 \times 256$ alternative model -- 
    instead of the default $(n_\mathrm{CNNl} = 128)\!:(N_\mathrm{dense} = 15) \times (n_\mathrm{dense} = 128)$ ResNet model, we show how well a $256\!:5 \times 256$ model performs for the \m87 training data in comparison, using the same hyperparameters.
    \item model variability -- prompted by intrinsic model variability being the dominating source of noise in our synthetic data \citep{eht-paperVI, 2022Satapathy, eht-SgrAv, zingularity1}, we have tested if our model overfits to the time-evolution of our training data. We have turned off the random shuffling of \m87 training data and used a $\eta_\mathrm{val} = 0.5$ split across the number of \ac{grmhd}-\ac{grrt} image frames for each model. The training was done with synthetic data generated from the first 50\,\% of model frames and validation with the latter 50\,\%. Here, we have used 450,000 training samples.
    \item model variability control -- as a control study and baseline for the model variability, we have done the same as above but with the random shuffling enabled again.
    \item fiducial \sgra model -- we show the training performance metrics for our \ac{eht} 2017 \sgra model described in \Cref{sec:eht_grmhd_grrt_network_implement}.
    \item \sgra $1024\!:6 \times 64$ alternative model -- we show how well an alternative $n_\mathrm{CNNl} = 1024, N_\mathrm{dense} = 6, n_\mathrm{dense} = 64$ model performs for the \sgra training data.
    \item \sgra Stokes $\mathcal{I}$ -- we show the performance of our fiducial \sgra model when trained on only Stokes~$\mathcal{I}$ data instead of the full polarization information content from all correlation products. Here, we have used 30,000 training samples.
    \item \sgra thermal-noise-only data -- instead of doing the full \symba forward modeling for $\widetilde{S}$ \citep{zingularity1}, we have created 25,000 \sgra training samples with only thermal noise added as $\widetilde{C}$.
\end{enumerate}

The training of our fiducial BANNs has converged with low error rates. Most noticeably, the MAD/SANE magnetic states are easily distinguishable for the standard sets of models in our classifiers. The performances of our BANNs are unaffected by small changes in the networks' architectures and hyperparameters.
We see smaller validation errors for \m87 compared to \sgra, which is most likely due to a combination of two effects. Firstly, we have $2.4$ times more training data for \m87. Secondly, the intrinsic model variability of \sgra within a single training sample translates into additional noise for the parameter inference.
For the \m87 models, $R_\mathrm{high}$ shows the highest validation errors, probably because $R_\mathrm{high}$ has not much influence on the GRRT image morphology for MAD models \citep{eht-paperV}.

The model variability tests show that our network is able to generalize well, as our control study has the same magnitude of validation errors. 
The robust data features used for the model parameter discrimination are not bound to the particular image snapshot realizations of the time-variable source structure. We note that both the \sgra and \m87 models show substantial intrinsic time variability, but on different time scales. While \sgra varies within \ac{eht} observing tracks, \m87 allows for a cleaner variability study with a clear cut across single model frames per observing track.
Given that our fiducial model performs substantially better than our variability studies, we see that a sufficiently large training dataset of several hundred thousand samples is needed to bring down BANN errors to low levels.

Without the polarization information and with only a few $N_\mathrm{tr}$, our \sgra models perform poorly and are barely being trained at all. The polarization has the biggest impact on the MAD/SANE classification, which is evident from the direct relation to the magnetic field structure. For spin measurements, a recent analysis by \citet{2023Ricarte} has shown the importance of the linear polarization structure.

When only thermal noise is added as $\widetilde{C}$, the validation errors are strongly reduced, even when only a few $N_\mathrm{tr}$ are used.
A neural network trained on synthetic data with lacking noise properties will thus likely overfit on observational data.
\vspace{0.15cm}
\subsection{Network hyperparameter survey}
\label{sec:zingularitysurvey}

Small parameter surveys have been performed to find the fiducial hyper-parameters for \m87 and \sgra listed in \Cref{tab:zingularity}. We surveyed 
\begin{itemize}[topsep=2ex, itemsep=1ex,partopsep=1ex,parsep=1ex]
    \item $\eta_\mathrm{drop} = (0,\, 0.01,\, 0.02)$,
    \item $\mathcal{L}_1 \;\;\,= (0,\, 0.01,\, 0.02)$,
    \item $\mathcal{L}_2 \;\;\,= (0,\, 0.01,\, 0.02)$.
    \item $N_\mathrm{ep} \;\,= (50,\, 60,\, 75,\, 100,\, 200)$.
\end{itemize}
We ran every parameter combination thrice for each source with different seed values for the BANN initialization at the start of training.
We identify networks as viable and stable when the loss is low and the three differently instantiated networks agree within 10\,\% for all inferred parameters. The parameter inference is computed for many bootstrapping realizations from the observational \m87 and \sgra $\widetilde{U}$ EHT data plus a few randomly selected validation data samples.

As the bootstrapping $\widetilde{C}$ is already incorporated into $\widetilde{S}$ and by pre-selecting networks with low validation errors, our stability criterion for the validation data is almost always fulfilled by construction.
For $\widetilde{U}$, the hurdle for the network to robustly generalize is higher. Typically, we see consistency across hyperparameters and different viable network architectures with some outliers occurring for a single inferred parameter for a particular random seed. We have rejected networks where such stochastic outliers can appear.

\begin{figure*}
    \begin{center}
    \includegraphics[width=0.495\textwidth]{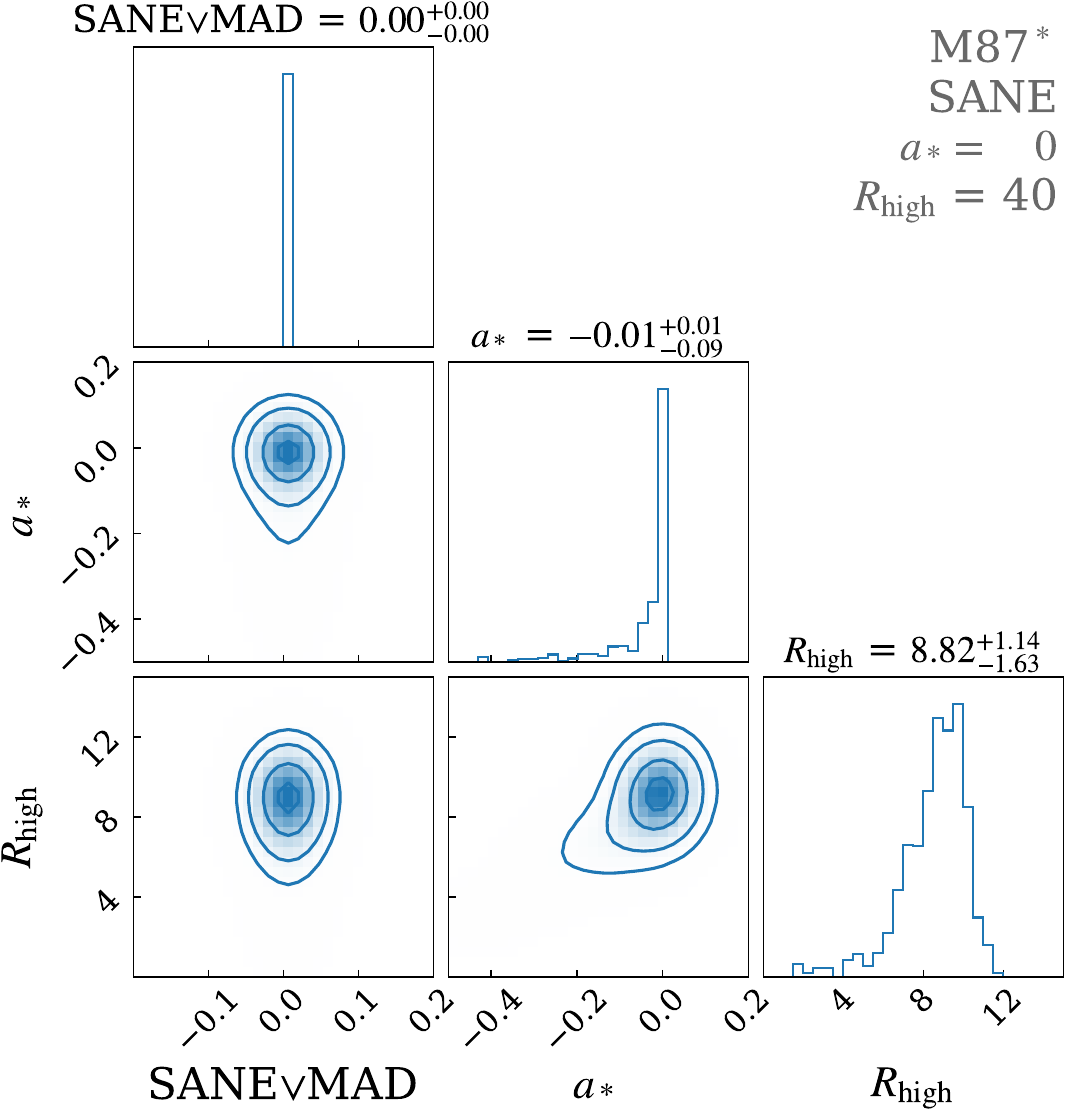}
    \includegraphics[width=0.495\textwidth]{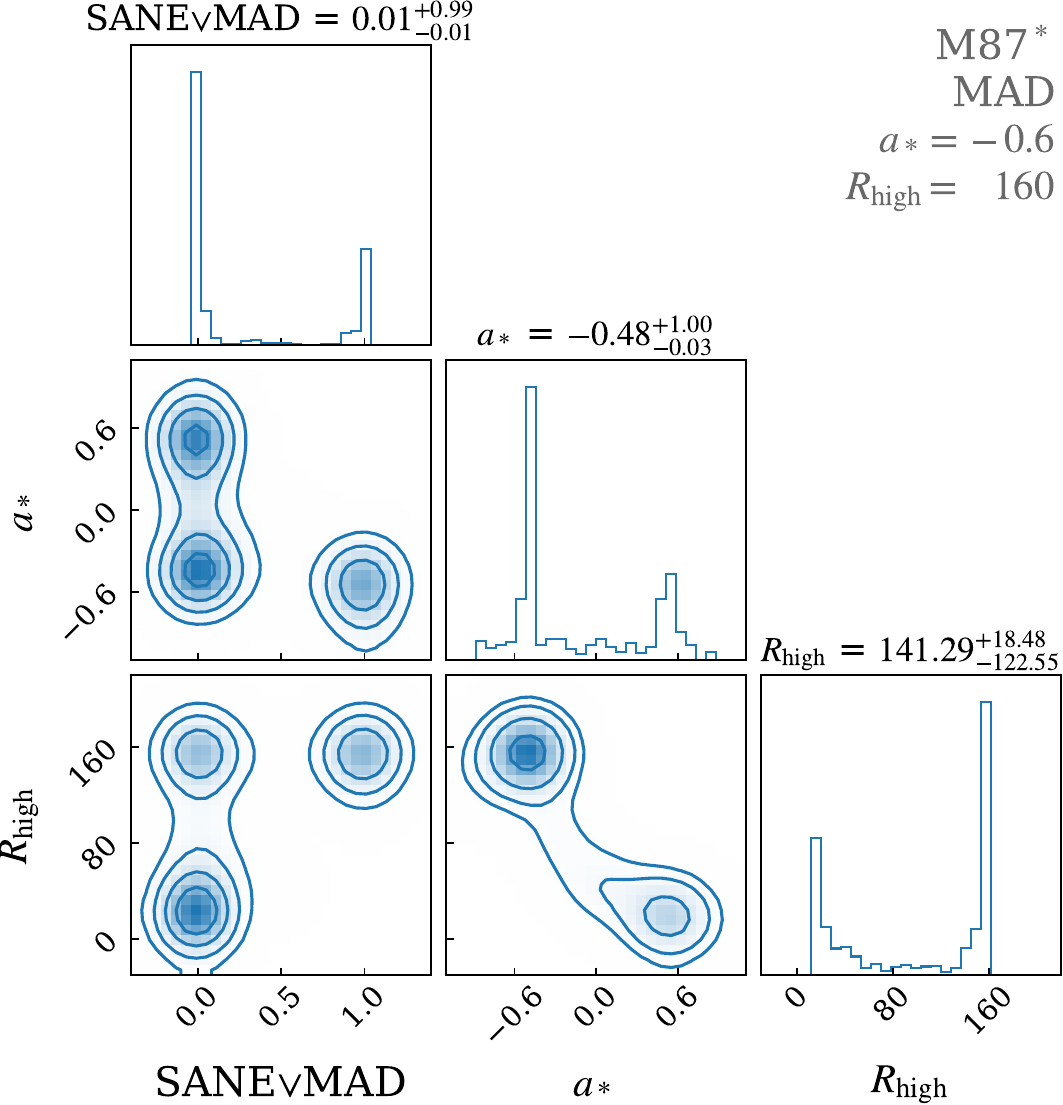}
    
    \vspace{0.2cm}
    \hrule
    \vspace{0.2cm}
    
    \includegraphics[width=0.495\textwidth]{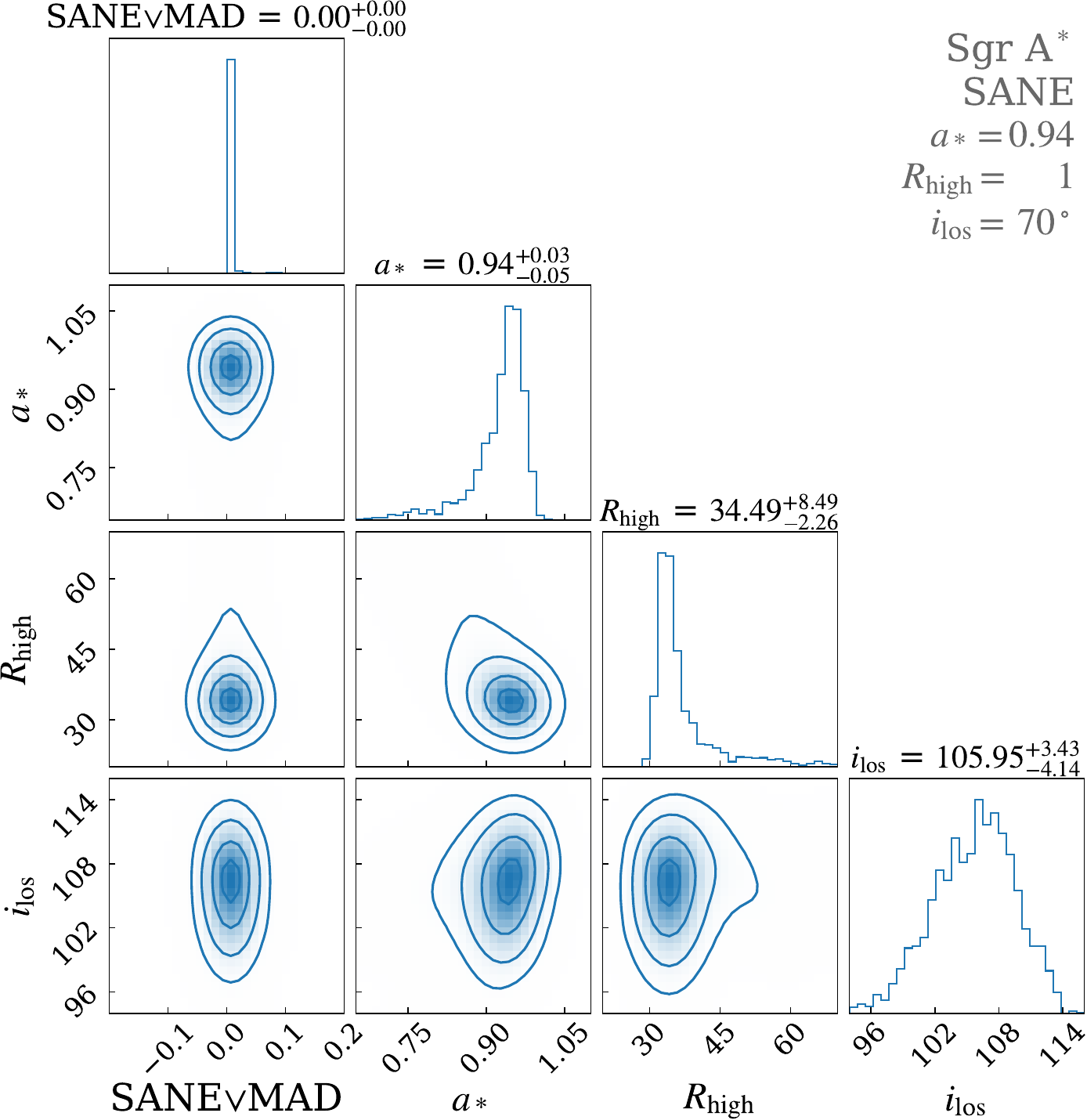}
    \includegraphics[width=0.495\textwidth]{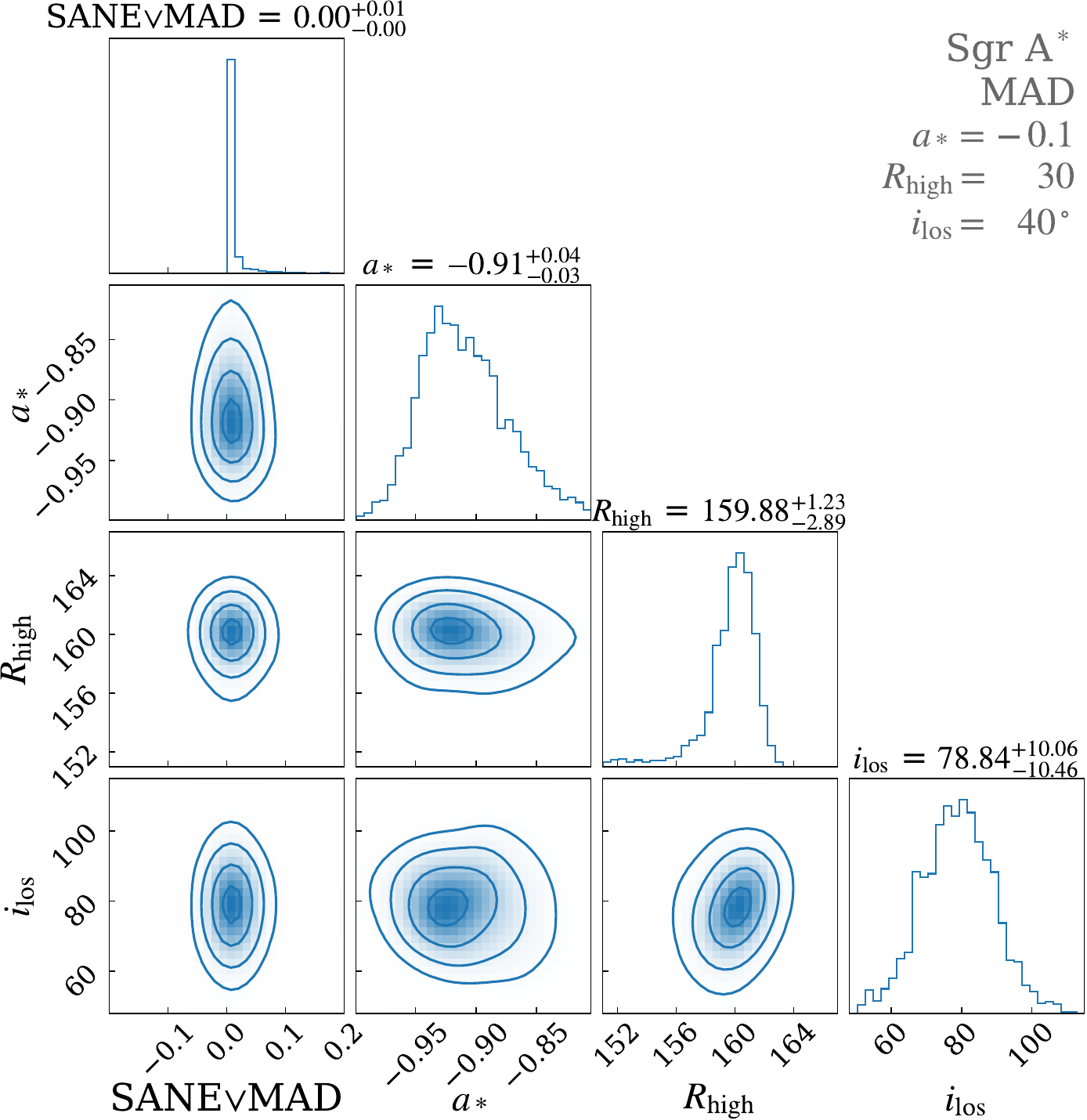}
    \caption{Inference results on \m87 (top row) and \sgra (bottom row) test datasets created from differing simulations (described in \Cref{sec:bootstrap_results}) with our fiducial BANN. The corner plots give the inferred parameters. Ground truth values are labeled in the top right corners. For the magnetic state, a value of zero corresponds to a certain SANE classification and a value of one to a certain MAD classification.}
    \label{fig:M87SGRAtestdata}
    \end{center}
\end{figure*}

\begin{figure*}
    \centering
    \includegraphics[width=0.495\textwidth]{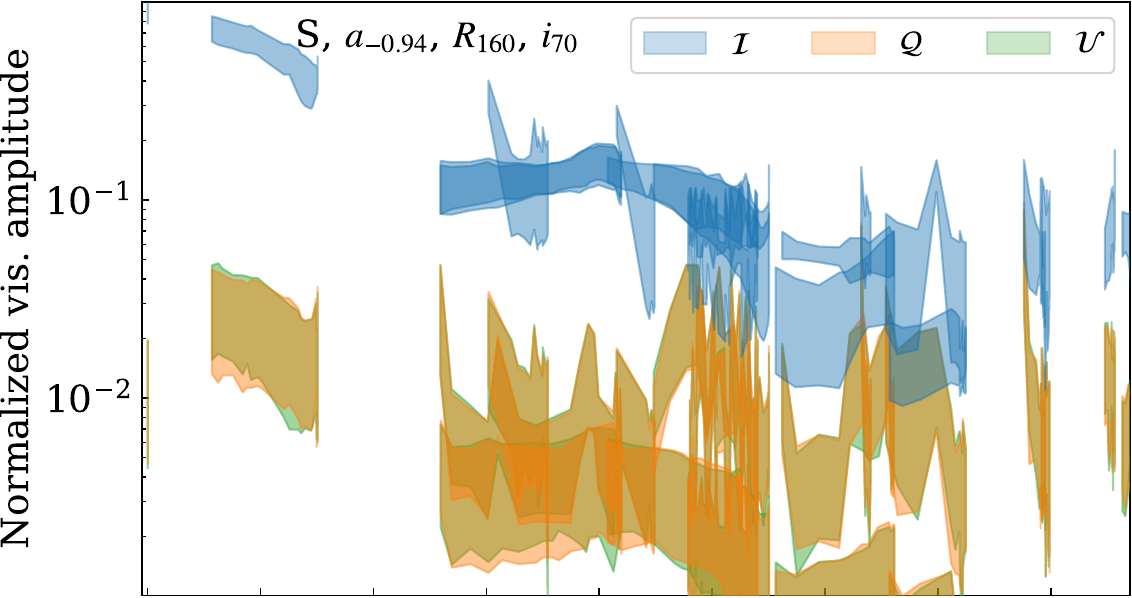}
    \includegraphics[width=0.495\textwidth]{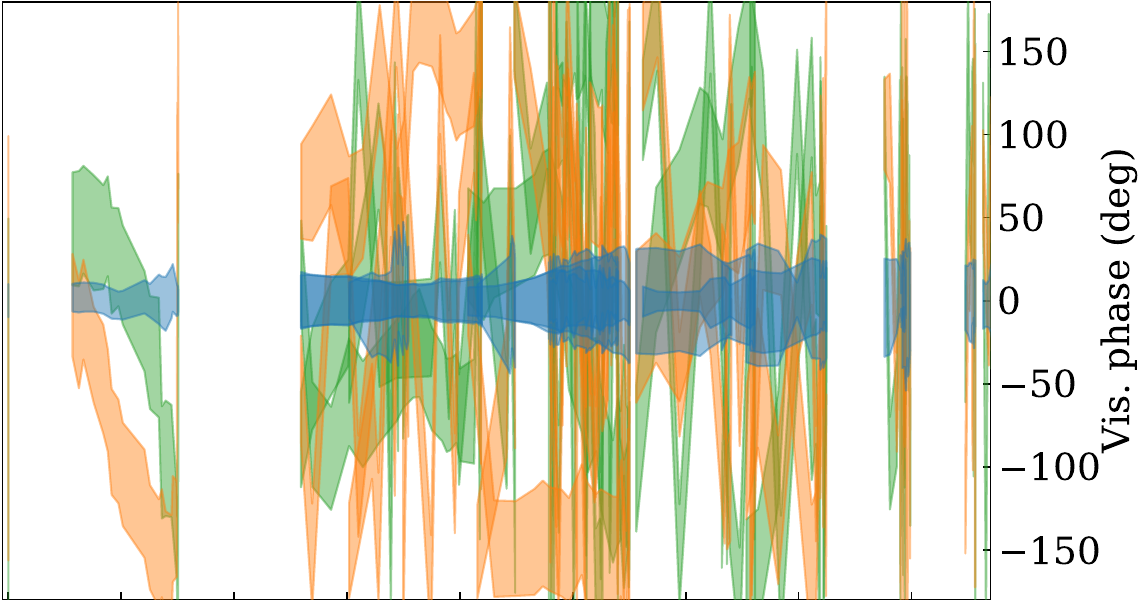}
    \includegraphics[width=0.495\textwidth]{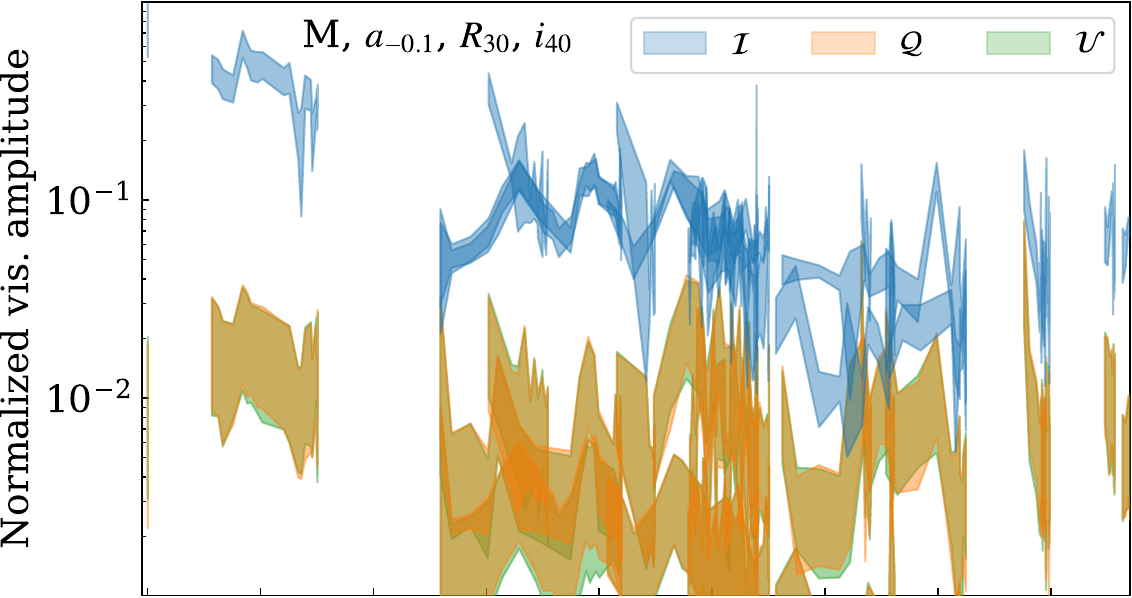}
    \includegraphics[width=0.495\textwidth]{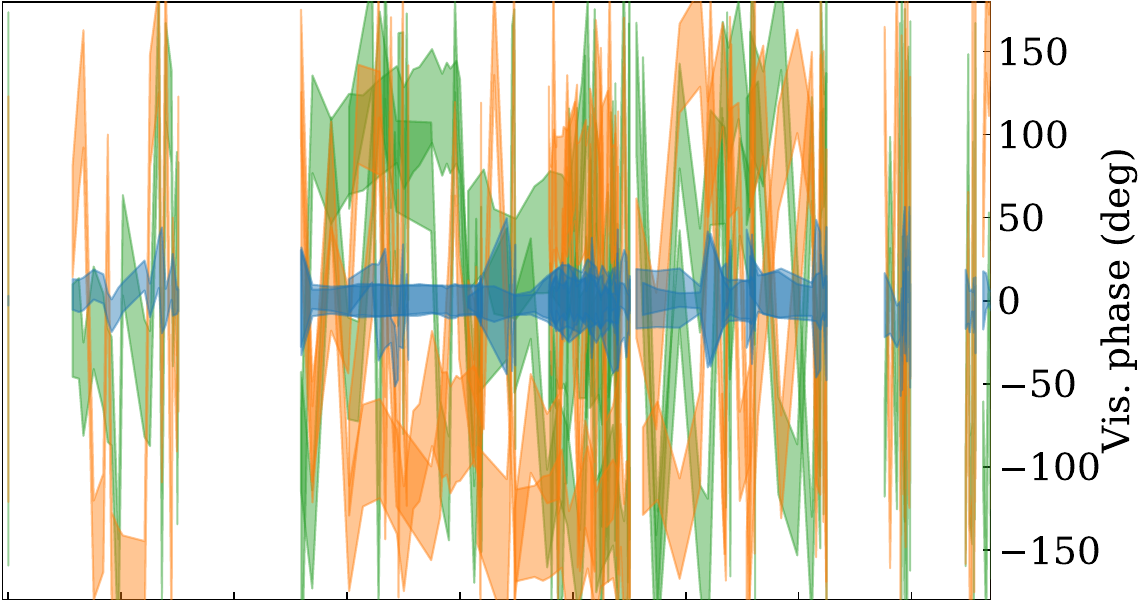}
    \includegraphics[width=0.495\textwidth]{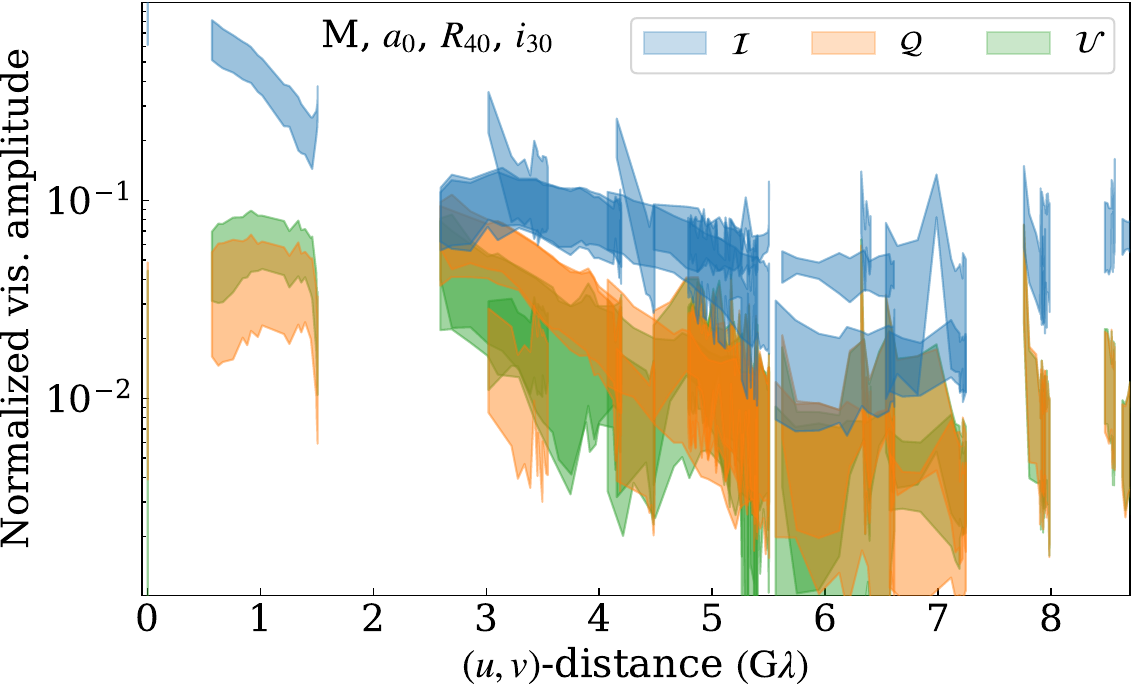}
    \includegraphics[width=0.495\textwidth]{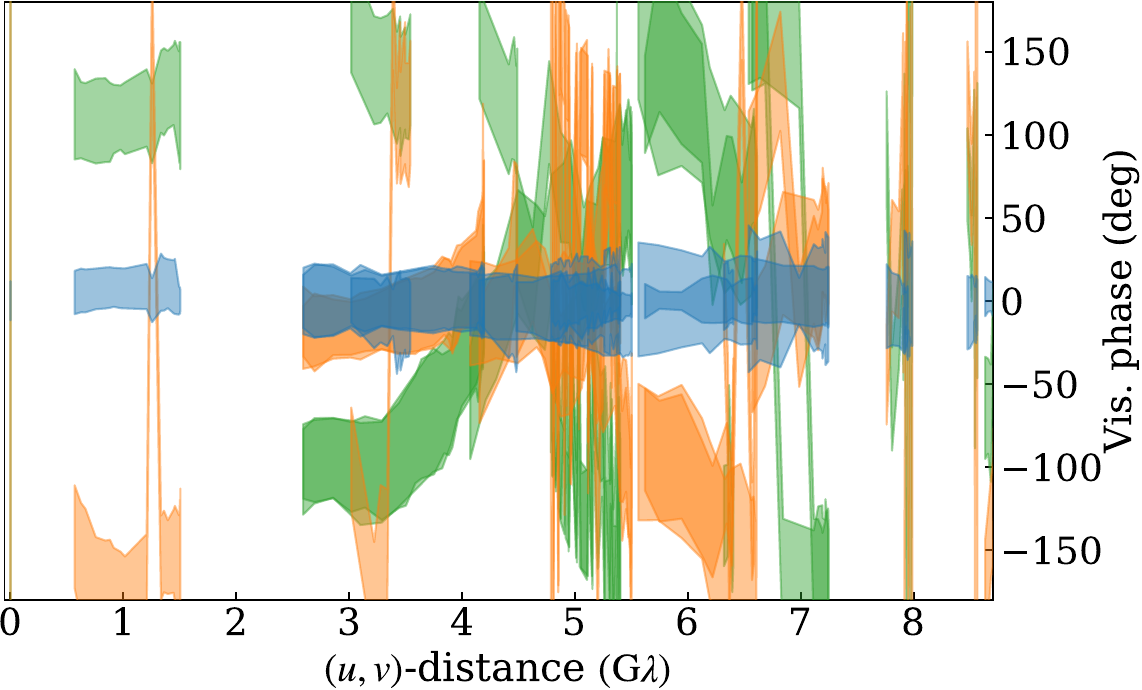}
    \caption{Normalized visibility amplitudes and phases in degrees (deg) color-coded by $\mathcal{I}$, $\mathcal{Q}$, $\mathcal{U}$ Stokes parameters with standard deviation error bands computed from 100 synthetic data realizations of 2017 \sgra EHT observations with corresponding \mbox{($u$, $v$)} coverage are displayed. The parameters for the three models considered here are given in the top left corners of the amplitude plots in the left column: a S(ANE) model in the top row and two different M(AD) models in the middle and bottom rows. The model in the middle is from test data, which has smaller error bars because the underlying simulation was run for a shorter time, leading to a smaller model variability.}
    \label{fig:modelcompare}
\end{figure*}
\vspace{0.15cm}
\subsection{Parameter inferences of test datasets}
\label{sec:bootstrap_results}

So far, we have established that our fiducial BANN models give reliable results also on data not seen during training. However, all synthetic observations used so far were based on the same type of \textsc{kharma} \ac{grmhd}-\ac{grrt} models \citep{2022Wong, 2021Prather}.
In this section, we describe inferences on test datasets that were obtained from a different kind of simulation model: GRMHD runs from the \textsc{bhac} code \citep{Porth2017} ray-traced by \textsc{raptor} \citep{raptor1, raptor2}.
The consistency between our different codes has been established, but we are using different setups and assumptions in our different code libraries.
As explored in \citet{eht-SgrAv, 2024eht-sgra-pol}, \textsc{kharma} and \textsc{bhac} models show significant differences in the standard EHT model scoring cuts and different assumptions on adiabatic indices during ray-tracing impact the electron temperature as well as polarimetric quantities.
Hence, synthetic data from \textsc{bhac}-\textsc{raptor} models can be used to test the robustness of our physical parameter inference with respect to nuisance variables in a comparison with the standard EHT model scoring.
As an additional check concerning the known issue of \ac{grmhd} model variability, we note that the \sgra test data model images are sampled with a cadence of 200\,s as opposed to the 100\,s of our standard models used for training and validation. Synthetic datasets are formed from movies of many frames over the course of a \sgra \ac{eht} \ac{vlbi} observing track.
The test data can be used to make a final model selection in case the parameter surveys leave multiple equally viable models.

We investigated the following models:
\begin{enumerate}[topsep=2ex, itemsep=1ex,partopsep=1ex,parsep=1ex]
    \item \m87, SANE, $a_*=0$, $R_\mathrm{high}=40$.
    \item \m87, MAD, $a_*=-0.6$, $R_\mathrm{high}=160$; $a_*$ falling outside of the grid of sampled parameters in the models used for the training data.
    \item \sgra, SANE, $a_*=0.94$, $R_\mathrm{high}=1$, $i_\mathrm{los}=\ang{70}$.
    \item \sgra, MAD, $a_*=-0.1$, $R_\mathrm{high}=30$, $i_\mathrm{los}=\ang{40}$; $a_*,\, R_\mathrm{high},\, i_\mathrm{los}$ falling outside of the training data grid.
\end{enumerate}
Parameter inferences are obtained with 1000 bootstrapping realizations times 1000 posterior draws.
The test results are shown in \Cref{fig:M87SGRAtestdata}.
The two left panels show models with parameters that fall within the training data grid. We ascribe the small $R_\mathrm{high}$ and $i_\mathrm{los}$ errors to the aforementioned differences in ray-tracing.

The \m87 $a_*=-0.6$ model shows a multimodal posterior as the network cannot unambiguously associate the data features with the closest matching models from the training data.
Even though the ground truth parameters are in the posterior, this test presents a failure mode of the network.
The \sgra MAD, $a_*=-0.1$, $R_\mathrm{high}=30$, $i_\mathrm{los}=\ang{40}$ model has been misidentified as SANE, $a_*=-0.9$, $R_\mathrm{high}=160$, $i_\mathrm{los}=\ang{80}$ model. The \Cref{fig:modelcompare} visibility data comparison for three \sgra models explains this discrepancy and elucidates the fiducial data features identified by the BANN. 
We denote the studied synthetic data as follows: \texttt{A} for the SANE, $a_*=-0.94$, $R_\mathrm{high}=160$, $i_\mathrm{los}=\ang{70}$ data, \texttt{B} for MAD, $a_*=-0.1$, $R_\mathrm{high}=30$, $i_\mathrm{los}=\ang{40}$, and \texttt{C} for MAD, $a_*=0$, $R_\mathrm{high}=40$, $i_\mathrm{los}=\ang{30}$.
\texttt{B} is our test data, \texttt{C} is training data with the closest matching parameters to \texttt{B}, and \texttt{A} is training data with the closest matching parameters to the posterior obtained from \texttt{B} (\Cref{fig:M87SGRAtestdata}).
Both \texttt{A} and \texttt{B} have matching $\mathcal{Q}$ and $\mathcal{U}$ amplitudes, while \texttt{C} shows clear offsets. The amplitude differences between $\mathcal{Q}$, $\mathcal{U}$ and $\mathcal{I}$ are also similar for \texttt{A} and \texttt{B}, but much smaller for \texttt{C}.
For the phases, $\mathcal{Q}$ and $\mathcal{U}$ are constant at the short baseline around 1\,G$\lambda$ for \texttt{C}, while \texttt{A} and \texttt{B} display variability.
Similarly, the $\mathcal{U}$ and particularly $\mathcal{Q}$ phases vary much more with baseline length for \texttt{A} and \texttt{B} compared to \texttt{C}.
All in all, it is not surprising that model \texttt{A} parameters are inferred from \texttt{B}, given the similarity of the visibilities.

Hence, the \texttt{B} misidentification is due to the limited grid of model parameters in the training data, which in turn is a result of the computational cost of \ac{grmhd} simulations.
This exercise highlights the model-dependence of our method. The inference results point to an interpolated parameter space of the \ac{grmhd}-\ac{grrt} training data models.
Note that the standard EHT model scoring suffers from the same problem, being restricted to the same type of sparsely sampled model libraries.
Ideally, we would have a finer model parameter grid to include data like \texttt{B} in $\widetilde{T}$. A better trained BANN could then either be sufficiently complex to be able to distinguish between \texttt{A} and \texttt{B} or the learned similarity between models would show up as uncertainty in the posterior, similar to the multimodal posterior of the \m87 $a_*=-0.6$ model.

Through our hyperparameter surveys and test data inferences, we have identified the best BANNs for \sgra and \m87. In \citet{zingularity3}, we will use these fiducial models for inference rather than averaging results over many models that would be acceptable but have worse performance.
As opposed to the very small validation errors from the training diagnostics, the posteriors shown in this section give more realistic indications of the true inference uncertainties of our BANNs, which in the end depend on the specific input data $\widetilde{U}$.

\section{Summary and conclusions}
\label{sec:zingularityconclusions}

In this second manuscript from a series, we presented our open-source \z{} framework for computationally efficient BANN training, validation, and inference.
As a first \z{} application, we used a comprehensive \ac{grmhd}-\ac{grrt} library of synthetic mm~\ac{vlbi} observations, to study how well physical parameters of the \sgra and \m87 AGN systems can be constrained with observations of the \ac{eht}.
We described our end-to-end pipeline, from the theoretical source models and observational data to the final posteriors. For scientific reproducibility, we made use of an open-source workflow management system and containerized versions of our synthetic data generation and machine learning software.

Compared to similar studies in the literature, we included a wider range of theoretical models and ensured that the intrinsic model variability was properly managed.
From these models, we created a comprehensive library of mock \ac{eht} observations. We considered a wide range of data corruption effects, both in the realistic synthetic data generation and the bootstrapping of uncertainties in the parameter inference process. Furthermore, we circumvented coherence losses and influences of data calibration methods on the measurements. The influence of coherence losses in the \ac{eht} data is described in Section~3.3 of \citet{eht-SgrAii}, for example. Here, we could use the full-Stokes information content of visibilities averaged in short (10\,s) time bins with small computational costs, thanks to efficient algorithms and data formats within the \tensorflow{} framework.
Influences from the use of particular data reduction strategies are visible in the geometric modeling results of the April 6, 2017 \sgra \ac{eht} data shown in Figure~30 of \citet{eht-SgrAiv}, for example. Depending on the chosen modeling parameters, the fits converged to different values for the ring diameter, asymmetry, and position angle for the same data when it is calibrated with different methods.
Here, we incorporated the same, \citep[upgraded,][]{zingularity1} calibration process in the training and observational data, thus ensuring that calibration-specific biases on the visibilities were not erroneously being picked up as model-dependent features.

Through hyperparameter surveys and dedicated inference runs on test datasets, which also include actual observational data to bridge the synthetic gap, we were able to i) weed out BANN architectures that yield spurious results from problematic overfitting on the training data, ii) uncover the inner workings of our networks, and iii) identify shortcomings in the training data sampling and the associated uncertainties on our parameter inferences.
We found the training of our final selection of fiducial BANNs to be well converged with low validation errors and robust against variations in the network (hyper-)parameters. We can deal with the intrinsic variability of our models through our networks' ability to generalize well.

We demonstrated the importance of utilizing the full visibility data content for the BANN training -- the polarization information is essential for the GRMHD parameter inference, particularly for the MAD--SANE magnetic field configuration. Additionally, showed that a sufficiently large training dataset is needed to achieve low validation errors and that realistic forward modeling is required for the training data generation. When the multitude of additional data corruption processes affecting EHT data are ignored, artificially low errors are obtained. Thus, a network trained on data where only thermal noise is added would be sensitive to data corruption effects present in observational data, leading to incorrect parameter inferences.
Shortcomings in previous machine-learning-based EHT analyses can likely be explained by a combination of these three effects: not enough training data, not utilizing the full information content of the data, and not taking into account all relevant data corruption effects.

\section{Outlook}
\label{sec:zingularityoutlook}

The parameter posteriors obtained by fitting the trained BANNs to observational \ac{eht} data are presented in \citet{zingularity3}.
Beyond the models studied in this work, one could consider the effects of particle acceleration \citep[e.g.,][]{Dexter2012, davelaar2018, 2019Davelaar, 2020Davelaar, 2021Yao, 2021Chatterjee, 2022CruzOsorio, 2022Fromm, 2023Zhao}, alternatives to the $R_\mathrm{high}$ electron temperature prescription and heating plus cooling effects \citep[e.g.,][]{2012Dibi, koral1, 2015Ressler, 2017Skadowski, 2018Chael, 2018Ryan, 2019Chael, 2020Anantua, 2020Yoon, 2021Mizuno, 2024Salas, 2024Moscibrodzka}, nonideal MHD and magnetic reconnection \citep[e.g.,][]{2019Ripperda,2020Ripperda,2021Chashkina, 2022Ripperda,2022Nathanail,2022Crinquand}, pair production \citep[e.g.,][]{2011Moscibrodzka, 2020Crinquand}, gas compositions other than pure hydrogen \citep[e.g.,][]{2022Wong_b}, tilted accretion disks \citep[e.g.,][]{2007Fragile,2013McKinney,2014MoralesTeixeira,2019White,2019Liska,2023Chatterjee}, different boundary conditions of the accretion flow \citep[e.g.,][]{2010Shcherbakov,2018Ressler,2023Olivares}, improved approximations for synchrotron radiative transfer calculations \citep{2024Davelaar}, and further non-Kerr models included in more advanced \ac{grmhd}-\ac{grrt} simulations in the future. Moreover, by employing frequency-resolved data, we could use spectral indices as well as rotation measures as discriminating model features, and would benefit from multifrequency synthesis \citep{1990Conway}. We also note the possibility of training on data from multiple observing days. Finally, the \z{} application to \ac{eht} data described here can easily be extended to AGN jets observed at larger scales \citep[e.g.,][]{2016Fromm,gena,2021MacDonald}. Incorporating jet emission and higher energy emission observations will lead to an improved inference, particularly for the electron distribution function.

\begin{acknowledgements}

We thank the anonymous referee for their insight and helpful
suggestions that have improved this paper.

This publication is part of the M2FINDERS project which has received funding from the European Research Council (ERC) under the European Union's Horizon 2020 Research and Innovation Programme (grant agreement No 101018682).

JD is supported by NASA through the NASA Hubble Fellowship grant HST-HF2-51552.001A, awarded by the Space Telescope Science Institute, which is operated by the Association of Universities for Research in Astronomy, Incorporated, under NASA contract NAS5-26555.

MW is supported by a Ram\'on y Cajal grant RYC2023-042988-I from the Spanish Ministry of Science and Innovation.

This material is based upon work supported by the National Science Foundation under Award Numbers DBI-0735191,  DBI-1265383, and DBI-1743442. URL: \url{www.cyverse.org}.
This research was done using resources provided by the Open Science Grid, which is supported by the National Science Foundation award \#2030508.
This research used the Pegasus Workflow Management Software funded by the National Science Foundation under grant \#1664162.
Computations were performed on the HPC system Cobra at the Max Planck Computing and Data Facility
This research made use of the high-performance computing Raven-GPU cluster of the Max Planck Computing and Data Facility.
Corner plots of posteriors were created with \texttt{corner.py} \citep{corner}.

\end{acknowledgements}

\bibliographystyle{aa} 
\bibliography{link_to_init}

\begin{appendix}

\section{Theory of a supervised Bayesian feedforward neural network}
\label{sec:NNbackground}

In a feedforward ANN, information flows only in one direction, from the first, to the second, to the third until the final output layer.
Here, we give a brief mathematical description of such networks.
We make use of the following naming conventions:
\begin{itemize}
    \item $M$ are the total numbers of layers in the network. Individual layers are numbered as $m = 0, 1, 2, ..., M-1$.
    Each layer consists of a number of neurons.
    \item $X_m$ is the output data of layer $m-1$ and input data of layer $m$. $X_0 \in \widetilde{T}$ is the input data of the network (i.e., a tensor of features that the network trains on). $X_M$ are the output predictions of the network.
    We denote the output of a specific neuron $j$ in the $m$th layer with $x_m^{(j)}$.
    \item $Y$ are the labeled targets of the training data $\widetilde{T}$ that will be compared against $X_M$ for supervised learning. These are integer representations when $X_0$ belongs to specific classes that are to be identified and/or real numbers for regression tasks of continuous variables of interest belonging to $X_0$.
    \item $L$ is the loss function used to compute the error $E$ between the target $Y$ and predicted $X_M$ outputs as $E = L(Y, X_M)$. Together with a learning rate $l_r$, this error is used to update (train) the weights of the network. 
    \item $f_m$ are activation functions used for layer $m$. These functions have to be nonlinear if the network is to learn nonlinear behavior in the input data $X_m$. In recent years, the rectified linear unit $f_m(z) = \mathrm{max}(z, 0)$ \citep[ReLU,][]{relu} has frequently been used.
\end{itemize}

Depending on the dimensionality of the data as it is passed through the network, $X_m$ can be a highly dimensional tensor.
With the above conventions, the output of layer $m$ of an ANN can be written as
\begin{equation}
X_{m+1} = f_m\left( W_m \cdot X_m + B_m \right)\,.
\end{equation}

Here, $W_m$ and $B_m$ are the tunable weight and bias tensors of layer $m$ that are being fitted while the network is being trained.
The weights are multiplied with the output of the previous layer and therefore depend on the dimensionality of $X_m$ and the specified output dimension $D_m \equiv \mathrm{dim}(X_{m+1})$. Single bias terms are added to the output of each neuron. The dimensionality of $B$ therefore depends only on $D_m$.
We denote $W_m \cdot X_m + B_m$ as the weighted input $Z_m$ of the layer $m$. The input to a specific neuron $j$ will be written as $z_m^{(j)}$.

We can now write the ANN algorithm in functional form as 
\begin{widetext}
\[
    \mathcal{F}(\{W\}, \{B\};\,X_0)
    \\ =
    f_{M-1} \left( W_{M-1} \cdot f_{M-2} \left( W_{M-2} \cdot f_{M-3} \left( \cdot \cdot \cdot f_1 \left( W_1 \cdot f_0\left( W_0 \cdot X_0 + B_0 \right) + B_1 \right) + ... + B_{M-3}\right) + B_{M-2}\right) + B_{M-1} \right)\,.
\]
\end{widetext}

Here, $\{W\}$ and $\{B\}$ denote the sets of all weights and biases that we will be optimizing, respectively. We denote individual weights of the connection between the $k^\mathrm{th}$ neuron in the $(m-1)^\mathrm{th}$ layer to the $j^\mathrm{th}$ neuron in the $m^\mathrm{th}$ layer as $w_m^{(jk)}$.
Similarly, $b_m^{(j)}$ will be the bias of the $j^\mathrm{th}$ neuron in the $m^\mathrm{th}$ layer.
Two commonly used types of layers, are 1) fully connected layers, where each neuron $j$ in the layer is connected to each neuron $k$ in the previous layer and 2) convolution layers \citep{cnn1, cnn2}, which can be seen as having a small receptive field where most entries of $W$ are zero: each neuron $j$ is only connected to a few neurons in the previous layer. Convolution layers form the basis of convolutional neural networks (CNNs).

For multidimensional prediction problems, individual parallel output layers are commonly used for each individual regression and classification task.
All of these output layers are connected to the same second to last layer in a feedforward ANN.
Typically, linear activation functions are used for regression and the softmax function $\sigma_s$ is used for classification.
Given a vector $\mathbf{s} = (s_1, s_2, ..., s_{N_\mathrm{c}})$ for $N_\mathrm{c}$ different classes, $\sigma_s: \mathbb{R}^{N_\mathrm{c}} \rightarrow [0, 1]^{N_\mathrm{c}}$ computes the probability of each class as
\begin{equation}
    \sigma_s(\mathbf{s}_i) = \frac{e^{s_i}}{\sum_{j=1}^{N_\mathrm{c}}e^{s_j}}\;\;\; \mathrm{for}\;\;i=1, 2, ..., N_\mathrm{c}\,.
\end{equation}

For multidimensional prediction problems, the target labels $Y$ have to be normalized to ensure an equal weighting for all predictions in the loss $L$.

Below, we describe the backpropagation algorithm, which is widely used to update $\left\{w_m^{(jk)}\right\}$ and $\left\{b_m^{(j)}\right\}$ during the training of ANNs.

Initially, $\{W\}$ and $\{B\}$ are set based on some a priori assumptions or randomly. In a forward pass through the network, $\mathcal{F}$ is computed to determine $L(Y, X_M)$.

For a neuron $j$ of the last layer, we can compute
\begin{equation}
    \delta_{M-1}^{(j)} = \frac{\partial L}{\partial x_{M-1}^{(j)}} f_{M-1}'\left(z_{M-1}^{(j)}\right)\,.
\end{equation}

Here, we use the Lagrange prime ($'$) notation for a derivative.
Subsequently, we can form the equivalent quantity for the complete last layer:
\begin{equation}
    \delta_{M-1} = \nabla_X(L) \cdot f_{M-1}' \left(Z_{M-1}\right)\,.
\end{equation}

Starting with the last layer, we can propagate backwards through the network layer by layer to compute
\begin{equation}
    \delta_m = f_m'\left(Z_m\right) \cdot W_{m+1} \cdot \delta_{m+1}
\end{equation}
for $m = M-2, M-3, ..., 0$. The individual $\delta_m^{(j)}$ are computed in the same way. Using a learning rate $l_r \leq 1$ to control the step size, each individual weight and bias parameters are then updated with
\begin{eqnarray}
    l_r \frac{\partial L}{\partial w_m^{(jk)}}&=&l_r x_{m-1}^{(k)} \delta_m^{(j)}\,,\\
    l_r \frac{\partial L}{\partial b_m^{(j)}}&=&l_r \delta_m^{(j)}\,.
\end{eqnarray}

In practice, a stochastic gradient descent is used for computational speed.
Instead of computing $\mathcal{F}$ and $L$ for the full dataset, we train on batches of size $N_\mathrm{b}$ at a time.

\subsection{Bayesian variational inference}

Starting with Bayes' theorem, the probability $P$ as a measure of belief for a set of neural network parameters $H$ given data $X$ can be written as
\begin{equation}
    P(H|X) = \frac{P(X|H) P(H)}{P(X)} = \frac{P(X|H) P(H)}{\int_H P(X|h)P(h){\mathrm d}h}\,.
\end{equation}

Here, $P(X|H)$, $P(H)$, $P(X)$, $P(H|X)$ are the likelihood, prior, evidence, and posterior, respectively.
The difficulty in sampling the posterior arises from the computational cost of evaluating the evidence integral.
Following \citet{BayesianNN}, we will now describe the variational inference method that is implemented in the Bayesian ANN (BANN) framework \z{} and used to sample from a surrogate variational distribution $q_\varphi(H) \sim P(H|X)$ instead of the exact posterior.

For non-Bayesian ANNs, $w_m^{(jk)}$ and $b_m^{(j)}$ are trained as single numbers. As such, $\mathcal{F}$ provides point estimates for inference and forward passes during training.
Bayesian networks fit trainable distributions for $w_m^{(jk)}$ and $b_m^{(j)}$ \citep{BayesianNN1, BayesianNN2, BayesianNN3, BayesianNN4}. For inference and forward passes, values are then drawn stochastically from the weight and bias posteriors.
The variational distribution is given by the combined distributions of all layers, parameterized by $\varphi$.
During training, $\varphi$ are optimized by minimizing the Kullback–Leibler divergence $D_\mathrm{KL}$ \citep{KL_Divergence}, such that $q_\varphi(H)$ approximates $ P(H|X)$ as closely as possible:
\begin{equation}
    D_\mathrm{KL}(q_\varphi||P) = \int_H q_\varphi(h)\log{\left(\frac{q_\varphi(h)}{P(h|X)}\right)}{\mathrm d}h \,.
    \label{eq:KLdivergence}
\end{equation}

Computing $D_\mathrm{KL}$ directly is expensive due to the integral over $P(h|X)$. Instead the evidence lower bound $\mathcal{E}$ is commonly evaluated:
\begin{eqnarray}
    \mathcal{E}_X(q_\varphi||P) &\equiv& \int_H q_\varphi(h)\log{\left(\frac{P(h,X)}{q_\varphi(h)}\right)}{\mathrm d}h \label{eq:elbo} \\  &=& \log{(P(X))} - D_\mathrm{KL}(q_\varphi||P) \,.
\end{eqnarray}

Here, $P(h,X)$ is the joint probability of $h$ and $X$. As the evidence $P(X)$ does not depend on $q$, maximizing $\mathcal{E}_X(q_\varphi||P)$ is equivalent to minimizing $ D_\mathrm{KL}(q_\varphi||P)$.

\end{appendix}
\end{document}